\documentclass[prx,aps,showpacs,twocolumn,amsmath,amssymb,floatfix,superscriptaddress]{revtex4-1}

\usepackage{hyperref}
\usepackage{graphicx}
\usepackage{bm}
\usepackage{color}
\usepackage[applemac]{inputenc}
\usepackage{amssymb}
\usepackage{tabularx}

\begin{document}

\title{Improved effective equation for the Rashba spin-orbit coupling in semiconductor nanowires}

\affiliation{Departamento de F{\'i}sica de la Materia Condensada,}
\affiliation{Departamento de F{\'i}sica Te{\'o}rica de la Materia Condensada,}
\affiliation{Condensed Matter Physics Center (IFIMAC) and Instituto Nicol\'as  Cabrera,
 Universidad Aut{\'o}noma de Madrid, E-28049 Madrid, Spain}

\author{Samuel D. Escribano}
\affiliation{Departamento de F{\'i}sica de la Materia Condensada,}
\affiliation{Condensed Matter Physics Center (IFIMAC) and Instituto Nicol\'as  Cabrera,
 Universidad Aut{\'o}noma de Madrid, E-28049 Madrid, Spain} 
\author{Alfredo Levy Yeyati}
\affiliation{Departamento de F{\'i}sica Te{\'o}rica de la Materia Condensada,}
\affiliation{Condensed Matter Physics Center (IFIMAC) and Instituto Nicol\'as  Cabrera,
 Universidad Aut{\'o}noma de Madrid, E-28049 Madrid, Spain}
\author{Elsa Prada}
\email[Corresponding author: ]{elsa.prada@uam.es}
\affiliation{Departamento de F{\'i}sica de la Materia Condensada,}
\affiliation{Condensed Matter Physics Center (IFIMAC) and Instituto Nicol\'as  Cabrera,
 Universidad Aut{\'o}noma de Madrid, E-28049 Madrid, Spain} 


 
\date{\today}

\begin{abstract}

Semiconductor Rashba nanowires are quasi-one dimensional systems that have large spin-orbit (SO) coupling arising from a broken inversion symmetry due to an external electric field. There exist parametrized multiband models that can describe accurately this effect. However, simplified single band models are highly desirable to study geometries of recent experimental interest, since they may allow to incorporate the effects of the low dimensionality and the nanowire electrostatic environment at a reduced computational cost. Commonly used conduction band approximations, valid for bulk materials, greatly underestimate the SO coupling in Zinc-blende crystal structures and overestimate it for Wurtzite ones when applied to finite cross-section wires, where confinement effects turn out to play an important role. We demonstrate here that an effective equation for the linear Rashba SO coupling of the semiconductor conduction band can reproduce the behavior of more sophisticated eight-band k$\cdot$p model calculations. This is achieved by adjusting a single effective parameter that depends on the nanowire crystal structure and its chemical composition.
We further compare our results to the Rashba coupling extracted from magnetoconductance measurements in several experiments on InAs and InSb nanowires, finding excellent agreement. This approach may be relevant in systems where Rashba coupling is known to play a major role, such as in spintronic devices or Majorana nanowires.
\end{abstract}

\maketitle

\section{Introduction}

The spin-orbit (SO) interaction is a relativistic effect that couples the electron's spin and momentum in the presence of an electric field. Among crystalline solids it is particularly strong in some semiconductors \cite{Winkler:03} and, although it is typically small compared to other characteristic energies, it produces a splitting of otherwise degenerate energy bands around the Fermi level \cite{Bychkov:JEPT84}. This can have tremendous consequences in the transport of electrons, as is manifested in the field of spintronics \cite{Zutic:RMP04} and, more recently, in spin-orbitronics \cite{Manchon:Nat15}. In particular, the SO interaction is the driving mechanism behind the existence of topological insulators \cite{Hasan:RMP10, Qi:RMP11} through the so-called quantum spin Hall effect \cite{Kane:PRL95}. It is also essential in the search of Majorana zero modes in topological superconductors \cite{Leijnse:IOP12, Alicea:IOP12, Beenakker:AR13, Sato:JPSJ16, Aguado:rnc17, Sato:IOP17}, such as the ones based on hybrid superconductor-semiconductor nanowires \cite{Lutchyn:PRL10,Oreg:PRL10}. In these wires, the SO term contributes to create non-degenerate bands with spin-momentum locking, a key ingredient behind the topological phase transition. Moreover, in the topological phase of the wire, the minigap that protects the Majorana modes from decoherence increases with the SO coupling \cite{Klinovaja:PRB12}.
 
For their connection to Majorana physics as well as for other spin-related mechanisms, semiconductor nanowires with strong SO coupling have come to the forefront of condensed-matter research \cite{Stanescu:IOP13,Lutchyn:NRM18,Prada:arxiv19}. There are several good reasons for their choice. They can be grown to a high degree of perfection, almost at the atomic scale \cite{Krogstrup:IOP13,Dasgupta:AM14,Barrigon:ChemRev19}. They can be proximitized \cite{Chang:Nnano15} both by depositing superconductors on top of them as well as by growing them epitaxially on the nanowire, forming well controlled and sophisticated heterostructures \cite{Krogstrup:Nat15}. They can also be easily contacted with metallic leads to an external circuit, and their properties are highly tunable through gate electrodes and external fields. In particular, this permits to tune them to enhance their SO coupling \cite{Liang:Nano12,Scherubl:PRB16,Takase:SciRep17}.
 
Inside these nanowires, electrons are subject to non-uniform electrostatic potentials. When a charged particle moves in an electric field, it experiences an effective magnetic field that couples to the particle's spin through the Zeeman effect \cite{Bychkov:JEPT84}.
The corresponding Hamiltonian is usually written as 
\begin{equation}
H_{\rm{SO}}=\vec{\alpha}\cdot (\vec{\sigma} \times \vec{k}),
\label{Eq:H_SO}
\end{equation}
where $\vec{k}$ is the electron's wavevector, $\vec{\sigma}$ is the vector of Pauli matrices in spin space and $\vec{\alpha}$ is the so-called SO coupling. 
This coefficient determines the strength of the coupling between the spin and the momentum of the electron and is related to the effective electric field felt by the electrons inside the wire \cite{Winkler:03}. Because the phenomena and applications mentioned before are very sensitive to the precise value of this coupling \cite{Manchon:Nat15}, a proper description of this mechanism is crucial to predict the actual properties of these nanowires.

There are two ways in which an electric field can arise in semiconductor nanowires \cite{Winkler:03}. On the one hand, the crystal itself creates an intrinsic electric field when there is a lack of an inversion centre (i.e., a bulk inversion asymmetry). This gives rise to the Dresselhaus SO coupling $\vec{\alpha}_{D}$ \cite{Dresselhaus:PR55}. On the other hand, an electric field arises when there is a lack of inversion symmetry due to an external confining potential (structural inversion asymmetry), due to e.g. crystal interfaces or metallic gates. This case corresponds to the Rashba coupling $\vec{\alpha}_{R}$ \cite{Rashba:59}. 
 
Depending on the source of the SO interaction, some theoretical methods may be more advantageous to describe it. Dresselhaus couplings, being an intrinsic interaction depending only on the crystal's unit cell structure and composition, are usually computed using \textit{ab initio} calculations \cite{Chantis:PRL06, Gmitra:PRB16}. Instead, Rashba couplings that depend on the electrostatic environment and/or interfaces with other materials are less amenable to \textit{ab initio} methods and tend to be described using effective models. In particular, multiband k$\cdot$p theory has been successfully used to compute the energy bands of Rashba semiconductors including several conduction and hole bands \cite{Winkler:03, Niquet:PRB06, Kishore:IOP12, Ehrhardt:14, Soluyanov:PRB16, Luo:AIP16, Campos:PRB18}.

Multiband effective models are specially suited for three-dimensional infinite systems whose bands depend on a single momentum $\hbar\vec{k}$. However, when treating low-dimensional systems (such as 2DEGs or nanowires), multiband models can be computationally challenging due to the extra degrees of freedom introduced by the transverse momentum subbands \cite{Ehrhardt:14}. This is because, for each transverse subband, one has to take into account several valence and conductions bands. This is further aggravated when treating heterojunctions with other materials \cite{Ehrhardt:14}. In this situation, it is very desirable to have a simplified effective theory that only takes into account the energy band of interest, typically the first conduction band, and that can incorporate the interaction with other less influential bands (such as heavy and light hole bands) through effective parameters.

For III-V binary compound semiconductors, which are broadly used in experiments, there exists an effective equation for the Rashba SO coupling based on a (spinful) single conduction band approximation, as we will show in Eq. (\ref{Eq:alpha_R}). This equation is derived from the so-called eight-band (8B) k$\cdot$p model, described in detail in Apps. \ref{App:A} and \ref{App:B}. Specifically, this equation takes into account the dependence of $\vec{\alpha}_R$ with $i)$ the electric field generated by a spatially-dependent electrostatic potential $\phi(\vec{r})$, $ii)$ the electron energy $E$, and $iii)$ the crystal structure and atomic composition through three effective parameters. One of these is known as the Kane parameter $P$ \cite{Winkler:03}, which represents the effective coupling between valence and conduction bands of the semiconductor. This equation, commonly used in the literature for its relative simplicity, is nevertheless the result of a lowest order expansion on the coupling between valence bands. This is a good approximation for bulk materials, but it fails considerably for structures where confinement is key, such as the nanowires that we study here.
As a result, the SO coupling obtained with this equation substantially differs from 8B model calculations for specific crystal structures. In this work we make the ansatz that the conduction band effective equation can accurately take into account different crystal structures and atomic compositions for quasi-1D nanowires if 
the Kane parameter is substituted by an improved one, which we call $P_{\rm{fit}}$. We calculate this parameter by fitting the SO coupling to 8B model results and provide its value in Table \ref{Table1} for InAs, InSb, GaAs and GaSb, both for Zinc-blende and Wurtzite structures. The \emph{improved} equation for the Rashba SO coupling constitutes the central result of this work and is given in Eq. \eqref{Eq:alpha_R_improved}, together with Table \ref{Table1}. It retains the simple functional form of the conduction band approximation, while incorporating the complexity of the crystal structure that appears due to confinement effects through one single parameter, $P_{\rm{fit}}$. We test its efficacy for different nanowire widths, different transverse subbands and for various electrostatic environments, with good results. We compare it with other popular simplified approximations which cannot capture the crystal structure complexity. Finally, we  demonstrate the reliability of our approach by contrasting our predictions with several experimental measurements of the Rashba coupling in InAs and InSb nanowires subject to different electrostatic environments, finding excellent agreement.

This paper is organized as follows. In Sec. \ref{Sec-Theory} we describe the type of materials of interest and the models and methods that we use. These comprise the 8B k$\cdot$p Kane model (Sec. \ref{Sec-8B_model}), the conduction band approximation (Sec. \ref{Sec-CB_approx}) including the improved effective equation for the SO coupling that we derive (Sec. \ref{Sec-CB_SO}), and the description of the electrostatic environment through the Schr\"odinger-Poisson equation and the Thomas-Fermi approximation (Sec. \ref{Sec-Electrostatic_environment}). In Sec. \ref{Sec-Results} we obtain the fitting $P_{\rm{fit}}$ parameter for various semiconductor compounds and crystal structures, collected in Table \ref{Table1}, and compare the SO coupling resulting from the different models and levels of approximation introduced before. In Sec. \ref{Sec-Applications} we compare our theoretical results with the Rashba coupling extracted from magnetoconductance measurements in five experiments on InAs and InSb nanowires, discussing the experiments in detail. Finally, in Sec. \ref{Sec-Conclusions} we present the conclusions of our work. This paper is complemented with several comprehensive appendixes on the 8B k$\cdot$p Kane model for Zinc-blende and Wurtzite  crystals (\ref{App:A}), the derivation of the conduction band approximation (\ref{App:B}), numerical methods (\ref{App:C} and \ref{App:D}), the comparison of the SO coupling between different semiconductor compounds (\ref{App:E}), the independence of $P_{\rm{fit}}$ with the electrostatic environment (\ref{App:F}) and the reliability of the conduction band approximation for the effective mass (\ref{App:G}).

\section{Models and methods}
\label{Sec-Theory}

\begin{figure}
\includegraphics[scale=1.33]{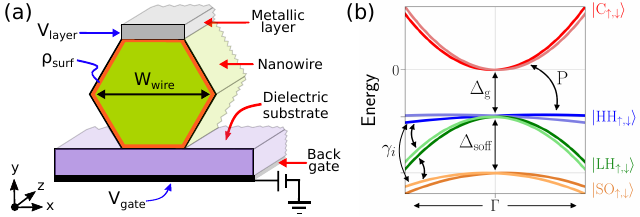}
\caption{(a) Sketch of the type of systems studied in this work: an infinite semiconductor nanowire with hexagonal cross section of width $W_{\rm{wire}}$ (green) is placed over a dielectric substrate (purple) and may be covered by a metal (grey) on one or several of its facets. A back gate (black) allows to tune the chemical potential inside the wire. At its facets, the nanowire may develop a charge accumulation layer that we simulate with a surface charge density $\rho_{\rm{surf}}$. This charge layer may be present or absent at the interfaces with metals, depending on the chemical details of the heterojunction. All these elements contribute to create an electrostatic profile inside the wire that in turn influences the Rashba spin-orbit (SO) coupling of its energy bands. (b) Schematics of the eight lowest-energy bands of III-V semiconductors around the $\Gamma$ point where these compounds exhibit a direct gap. These are grouped into four quasi-degenerate pairs that comprise the conduction band, the light-hole, heavy-hole and split-off valence bands. Their corresponding states, $|C_{\uparrow,\downarrow}\rangle$, $|HH_{\uparrow,\downarrow}\rangle$, $|LH_{\uparrow,\downarrow}\rangle$, $|SO_{\uparrow,\downarrow}\rangle$, serve as a truncated basis for the k$\cdot$p Kane model. $\Delta_g$ is the semiconductor gap between conduction and valence bands, $\Delta_{\rm{soff}}$ is the gap between valence and split-off bands, $P$ is the coupling between conduction and valence bands, and $\gamma_i$ are the intra-valence band couplings, see App. \ref{App:A}.}
\label{Fig1}
\end{figure}

As mentioned in the introduction, in this work we focus on III-V binary compound semiconductors such as InAs, InSb, GaAs, or GaSb, which typically exhibit a large Rashba SO coupling \cite{Vurgaftman:JAP01}. We consider semiconductor nanowires with crystal structures of Zinc-blende (111) or Wurtzite (0001), since they are the most commonly used in experiments due to their fabrication with low impurity concentrations \cite{Thelander:Nano10, Krogstrup:IOP13, Dasgupta:AM14}. We assume that the nanowires are infinite in the specific growth direction but with a finite hexagonal cross-section \cite{Fortuna:IOP10,Krogstrup:Nat15} of width $W_{\rm{wire}}$ as depicted in Fig. \ref{Fig1}. 

\subsection{Eight-band k$\cdot$p Kane model}
\label{Sec-8B_model}
Multiband k$\cdot$p models are known to successfully reproduce the energy-band structure of III-V compound semiconductors \cite{Winkler:03, Niquet:PRB06, Kishore:IOP12, Ehrhardt:14, Soluyanov:PRB16, Luo:AIP16, Campos:PRB18}. The SO coupling  can then be directly extracted from the shape of the energy spectrum \cite{Campos:PRB18,Moor:IOP18, Bommer:PRL19,Woods:PRB19}. We summarize here this procedure. 
These effective models, broadly explained and used in the literature (see e.g. Refs. \onlinecite{Winkler:03, Faria:PRB16}), assume that the electron movement through the crystal is well described by a single-particle Hamiltonian. This Hamiltonian includes relativistic SO effects as well as an effective potential, which arises due to electron-nuclei interactions and thus has the same periodicity than the Bravais lattice. This allows to use the Bloch theorem and to expand the periodic (Bloch) part of the wavefunction around a reciprocal-space point of interest. For the case of III-V semiconductors, the natural expansion is around the $\Gamma$ point where these compounds exhibit a direct gap \cite{Vurgaftman:JAP01}. The resulting Hamiltonian is then projected over a truncated basis set that includes explicitly the main bands of interest, while their couplings to the remaining bands are included perturbatively using L\"owdin perturbation theory \cite{Lowdin:ChemPhys51}. Some of the transition matrix elements are forbidden by crystal symmetries.
Using group theory arguments, the remaining couplings are substituted by effective parameters. These parameters are called Kane or Luttinger parameters and can be extracted from \textit{ab initio} calculations or experimental data for a particular material with a specific crystal structure. In the 8B model that concerns us, only the four (doubly quasi-degenerate) energy bands closer to the semiconductor gap are included in the basis set: the lowest-energy conduction band, and the heavy-hole, light-hole and split-off valence bands, see Fig. \ref{Fig1}(b). These eight bands are typically sufficient to account for the Rashba SO effects of these materials \cite{Luttinger:PR55,Kane:JPCS57,Faria:PRB16}. For a detailed derivation of this multiband k$\cdot$p Kane theory and the resulting 8B Hamiltonians for Zinc-blende and Wurtzite crystals, we refer the reader to App. \ref{App:A} and references therein \cite{ Darnhofer:PRB93, Bahder:PRB92, Winkler:03}. We note that these Hamiltonians are only accurate for the reciprocal space range in which they are fitted to \textit{ab initio} calculations. For the particular Kane parameters provided in App. \ref{App:A}, this implies $k\in [-1,1](nm)^{-1}$. Hence, when the Fermi wavevector is outside this range, the assumptions made for the k$\cdot$p approximation break down and this model is no longer reliable.

For finite cross-section nanowires that are infinite along the $z$-direction, the band structure is then calculated by diagonalizing the k$\cdot$p Hamiltonian for different $k_z$ values. To do so, we first replace the momentum operator components across the wire's section by their corresponding derivatives, i.e., $k_x\rightarrow -i\partial_{x}$ and $k_y\rightarrow -i\partial_{y}$, while considering $k_z$ as a good quantum number due to the translational invariance along this direction. Then, the operators are discretized using the finite difference method in a rectangular mesh for the nanowire section. Special treatment is required in this step in order to avoid spurious solutions in the energy spectrum \cite{Cartoixa:JAP03,Jiang:JAP14}. For an extended explanation of the numerical methods used in this work, see App. \ref{App:C}.

Once the band structure is obtained, the SO coupling can be extracted by fitting each subband $j$ by the following effective dispersion relation
\begin{equation}
E_{\pm}^{(j)}(k_z)=\frac{\hbar^2k_z^2}{2m_{\rm{eff}}^{(j)}}+E_T^{(j)}\pm\sqrt{(\alpha_{\rm{eff}}^{(j)}k_z)^2+(\beta_{\rm{eff}}^{(j)}k_z^2)^2},
\label{E_fit_kp}
\end{equation}
where $j$ and $\pm$ are the subband and spin indexes, $m_{\rm{eff}}^{(j)}$ is the effective mass and $E_T^{(j)}$ is the transverse subband energy at $k_z=0$. The other two parameters, $\alpha_{\rm{eff}}^{(j)}$ and $\beta_{\rm{eff}}^{(j)}$, take into account possible Rashba and/or Dresselhaus SO effects. While the Rashba contribution to the SO coupling is known to be mainly linear in $k_z$ irrespective of the crystal structure, the Dresselhaus one can be both linear and quadratic for Wurtzite crystals, and only linear for Zinc-blende (111) crystals \cite{Campos:PRB18}. Hence, it is not possible, in principle, to separate the contributions of the Rashba and Dresselhaus coefficients in the linear term of the SO coupling. In practice, the linear Dresselhaus contribution turns out to be zero for Zinc-blende (111) crystals and negligible or zero for Wurtzite (0001) ones \cite{Gmitra:PRB16, Faria:PRB16} \footnote{Notice that Zinc-blende (111) crystals are almost symmetric in the three directions and Wurtzite (0001) ones are almost symmetric around the z-axis. Therefore, since there is no bulk inversion asymmetry in these directions, there cannot be a SO interaction either. Only higher-order asymmetries, i.e., the ones accounted for by cubic terms in $k$ (although quadratic in $k_z$), can contribute to Dresselhaus terms. These terms are taken into account through the parameter $\beta_{\rm{eff}}$ in Eq. \eqref{E_fit_kp}.}. Thus we assume that the linear coefficient of the SO coupling $\alpha_{\rm{eff}}^{(j)}$ is dominated by the Rashba contribution. 

\subsection{Conduction band approximation}
\label{Sec-CB_approx}

We just outlined how to extract the SO coupling of a particular band starting from the 8B Hamiltonian that describes the crystal band structure and then applying a fitting procedure. This is certainly an indirect way that can become laborious and computationally expensive under certain circumstances, such as when dealing with low dimensional systems and, especially, when they form heterostructures with other materials. 
For this reason, we look for a simplified model that takes into account the band of interest, the conduction band in our case, but still captures the main SO effects as described by the 8B model. This is a standard procedure in the literature that leads to analytical expressions for the SO coupling under certain approximations (see App. \ref{App:B} for a full derivation).


Our starting point is thus the 8B Hamiltonian of a specific crystal, which we consider describes the system faithfully. We want to find an effective Hamiltonian within the conduction band, $H_{\rm{CB}}$, by integrating out the valence bands. Since this folding-down procedure cannot however be done exactly with a nonuniform electrostatic potential $\phi(\vec{r})$, it is customary \cite{Darnhofer:PRB93,Winkler:03} to ignore the couplings among the valence bands, $\gamma_i$  
(see Fig. \ref{Fig1}). This allows to
describe the conduction band with the following analytical \emph{reduced} Hamiltonian
\begin{eqnarray}
H^{(0)}_{\rm{CB}}=\vec{k}\frac{\hbar^2}{2m^{(0)}(\vec{r})}\vec{k}-e\phi(\vec{r}) \nonumber \\
+\frac{1}{2}\left[\vec{\alpha}_{R}^{(0)}(\vec{r})\cdot\left(\vec{\sigma}\times\vec{k}\right)+\left(\vec{\sigma}\times\vec{k}\right)\cdot\vec{\alpha}_{R}^{(0)}(\vec{r})\right],
\label{Eq:H_cb}
\end{eqnarray}
where $\vec{r}=(x,y,z)$ and we have chosen the conduction band edge at $\phi=0$ as the reference energy. This Hamiltonian is the zeroth-order term in an expansion of $H_{\rm{CB}}$ in $\hbar^2\vec{k}^2\gamma_i/(2 m_e \Delta_i)$, where $m_e$ is the bare electron mass and $\Delta_i$ represents the different energy gaps involved in the 8B Hamiltonian, see App. \ref{App:B}. In the previous expression, the effective mass coefficient is given by
\begin{eqnarray}
\frac{1}{m^{(0)}(\vec{r})}=\frac{1}{m_e}+ \frac{2P^2}{3\hbar^2}\left[\frac{2}{\Delta_g+e\phi(\vec{r})+E} \right. \nonumber \\
+ \left. \frac{1}{\Delta_{\rm{soff}}+\Delta_g+e\phi(\vec{r})+E}\right],
\label{Eq:m_eff}
\end{eqnarray}
and the linear Rashba SO coupling by
\begin{eqnarray}
\vec{\alpha}^{(0)}_{R} (\vec{r})=\frac{eP^2}{3}\left[\frac{1}{(\Delta_g+e\phi(\vec{r})+E)^2} \right. \nonumber \\
\left. -\frac{1}{(\Delta_{\rm{soff}}+\Delta_g+e\phi(\vec{r})+E)^2}\right] \vec{\nabla}\phi(\vec{r}).
\label{Eq:alpha_R}
\end{eqnarray}
Here, $\Delta_g$ and $\Delta_{\rm{soff}}$ are the semiconductor and split-off gaps, respectively, and $P$ is one of the Kane parameters of the 8B model that accounts for the coupling between conduction and valence bands, see Fig. \ref{Fig1}(b).
To find its expectation value, we need to project over the electron's wave function
$|\Psi\rangle$, $\vec{\alpha}^{(0)}_{R}=\langle\Psi|\vec{\alpha}^{(0)}_{R}(\vec{r})|\Psi\rangle$. 
Notice that, in the notation we are following, the gradient in Eq. \eqref{Eq:alpha_R} only acts on the electrostatic potential, i.e., it must not be applied on $|\Psi\rangle$. We further point out that Eqs. \eqref{Eq:H_cb}, \eqref{Eq:m_eff} and \eqref{Eq:alpha_R} are only applicable within the range $\Delta_g>-\left(e\phi(\vec{r})+E\right)$. This is because, outside this range, the conduction band and the first valence band at different spatial points in the wire can overlap in energy. Consequently, the system can no longer be described using just the conduction band approximation.

The previous expressions depend on the electron energy $E$, which is itself the solution to the eigenvalue problem of Hamiltonian of Eq. \eqref{Eq:H_cb}. To remove this dependence, which requires a self-consistent solution for the Hamiltonian eigenvalues, it is frequently assumed in the literature \cite{Darnhofer:PRB93, Winkler:03, Wojcik:PRB18} the more restrictive condition $\Delta_g\gg \left|e\phi(\vec{r})+E\right|$. Expanding the expressions above and truncating them to lowest order, the effective mass gets
\begin{eqnarray}
\frac{1}{m^{(0)}}\simeq\frac{1}{m_e}+ \frac{2P^2}{3\hbar^2}\left(\frac{2}{\Delta_g}+ \frac{1}{\Delta_{\rm{soff}}+\Delta_g}\right),
\label{Eq:m_eff-oversimplified}
\end{eqnarray}
and the SO coupling 
\begin{eqnarray}
\vec{\alpha}^{(0)}_R (\vec{r})\simeq\frac{eP^2}{3} \left[\frac{1}{\Delta_g^2}-\frac{1}{(\Delta_{\rm{soff}}+\Delta_g)^2}\right] \vec{\nabla}\phi(\vec{r}).
\label{Eq:alpha_R-oversimplified}
\end{eqnarray}
This last (\emph{simplified}) equation has been used in previous works \cite{Wojcik:PRB18, Winkler:PRB19, Escribano:PRB19} to describe the Rashba SO coupling in nanowires, in an attempt to go beyond the use of a rough constant coefficient. Nevertheless, as we will show in Sec. \ref{Sec-Results}, both this equation and the more general one Eq. \eqref{Eq:alpha_R} fail to predict the behavior of nanowires with specific Zinc-blende or Wurtzite crystal structures. The reason is that the intra-valence band couplings that have been ignored in the reduced Hamiltonian, although negligible for bulk crystals, turn out to be essential in low-dimensional structures where confinement effects are important.


\subsection{Improved SO coupling equation}
\label{Sec-CB_SO}


Going beyond the zeroth-order in the conduction band expansion of the SO coupling operator leads to complicated expressions, especially as the order increases, see App. \ref{App:B}. One can check numerically that summation over several terms is needed to approach the correct SO coupling. Since summing up the infinite series is not better than solving the 8B Hamiltonian, and since in this work we are looking for a manageable expression for the SO coupling, we resort to the following ansatz. We propose to use an expression for the SO coupling with the same functional form of Eq. \eqref{Eq:alpha_R}, which is the dominant term in the expansion, but where the parameter $P$ is substituted by an improved one, that we call $P_{\rm{fit}}$, chosen so as to reproduce the Rashba SO coupling extracted from the 8B model of Zinc-blende or Wurtzite nanowires. In this way, we are conjecturing that the lost information about the intra-valence band couplings $\gamma_i$ can be recovered, at least partially, by one fitting parameter. This parameter is going to depend on the semiconductor compound (InAs, InSb, etc.) and crystallography (Zinc-blende or Wurtzite). However, we assume that $P_{\rm{fit}}$ can be taken as independent of $\phi(\vec{r})$ and the electron's energy (or equivalently, $W_{\rm{wire}}$). Remarkably, as we will show in the rest of the paper, these assumptions turn out to be pretty accurate for realistic experimental conditions. 


We are interested in finite cross-section nanowires that are moreover translationally invariant along the $z$ direction. If the SO length is larger than the wire's diameter, i.e., $l_{\rm{SO}}\gtrsim W_{\rm{wire}}$, we can write the total energy as $E=E_T^{(j)}+E(k_z)$, where $E_T^{(j)}$ is the transverse subband energy and $E(k_z)$ the longitudinal part. For the small $k_z$-range for which the 8B model applies, the condition $|E_T^{(j)}|\gg |E(k_z)|$ is satisfied. Therefore, projecting over the transverse part of the Hamiltonian's eigenstates, see App. \ref{App:C}, we posit that we can write the $j$-th subband effective Rashba SO coupling as
\begin{eqnarray}
&&\vec{\alpha}^{(j)}_{R,\rm{improved}}=\left<\Psi_T^{(j)}\right|\frac{eP_{\rm{fit}}^2}{3} \left[\frac{1}{(\Delta_g+e\phi(x,y)+E_T^{(j)})^2} \right. \nonumber \\
&&\left. -\frac{1}{(\Delta_{\rm{soff}}+\Delta_g+e\phi(x,y)+E_T^{(j)})^2}\right]\vec{\nabla}\phi(x,y)\left|\Psi_T^{(j)}\right>,
\label{Eq:alpha_R_improved}
\end{eqnarray}
where $P_{\rm{fit}}$ is extracted by fitting this equation to the Rashba coupling obtained with the 8B Kane model. Note that in the previous expression the Rashba $z$ component is zero since the electrostatic potential depends only on the transverse coordinates. As before, the gradient only acts on $\phi$ and not on the wave function. This equation has to be solved self-consistently since $E_T^{(j)}$ is the energy associated to the transverse eigenstate $|\Psi_T^{(j)}\rangle$. We will show in Sec. \ref{Sec-Results} that this \emph{improved} equation produces results in good agreement with those of 8B model calculations at a considerably reduced computational cost. 

A similar improved equation could be proposed for the effective mass. However, this is not necessary as the zeroth-order conduction band approximation, Eq. \eqref{Eq:m_eff}, already produces very similar results to the 8B model ones (see Apps. \ref{App:B} and \ref{App:G}).


To finish this section, it should be noticed that the effects of confinement were also analyzed in Ref. \onlinecite{Winkler:03}, in particular for 2DEGs. Following a standard perturbative approach, the 8B Hamiltonian is first projected over a particular subband basis and then a folding-down procedure is performed. In this way, Winkler \emph{et al.} \cite{Winkler:03} arrive at an expression for the Rashba SO coupling [Eq. (6.23) in Ref. \onlinecite{Winkler:03}] written in terms of a sum over all conduction-valence subbands matrix elements. In our work, we follow a different approach and propose a heuristic expression for the SO coupling \emph{operator} where the complexity introduced by the valence subbands is taken into account (approximately) through the parameter $P_{\rm{fit}}$, resulting in a significant reduction in computational cost. To obtain the magnitude of the resulting SO coupling, we just need to project over the desired conduction subband. Our \emph{improved} equation is not the result of a perturbative calculation, but an ansatz for the resummation of all inter-subband processes. We will show in Sec. \ref{Sec-Results} that it produces results in very good agreement with those of 8B model calculations under rather general conditions. Moreover, thanks to this approach, we arrive to a practical/simple equation for the SO coupling in terms of the gradient of the electrostatic potential (instead of a long and system-specific sum of matrix elements), which is both manageable and helps to intuitively understand the physical origin of the SO interaction.

\subsection{Electrostatic environment}
\label{Sec-Electrostatic_environment}
The Rashba SO coupling given in Eq. \eqref{Eq:alpha_R_improved} depends on the gradient of the electrostatic potential. This is a direct manifestation that only when there is a structural inversion asymmetry, i.e., an inhomogeneous electrostatic potential, there is a non-zero Rashba SO coupling. Hence, the SO coupling is sensitive to the precise electrostatic environment, as well as to electron-electron interactions. 
To compute the electrostatic potential corresponding to an arbitrary environment we use the Poisson equation given by
\begin{equation}
\vec{\nabla}\cdot\left[\epsilon(\vec{r}) \vec{\nabla}\phi(\vec{r})\right]=\rho_T(\vec{r}),
\label{Eq:Poisson}
\end{equation}
Here $\epsilon(\vec{r})$ is the inhomogeneous dielectric permittivity, which we take as constant inside each material and with abrupt changes at the interfaces (therefore, it encodes the geometry information of the environment), $\phi(\vec{r})$ is the electrostatic potential in the entire system and $\rho_T$ is the total charge of the wire. This source term includes two parts \cite{Winkler:PRB19},
\begin{equation}
\rho_T(\vec{r})=\rho_{\rm{mobile}}(\vec{r})+\rho_{\rm{surf}}(\vec{r}).
\label{Eq:Charge_density}
\end{equation}
The first one represents the mobile charge inside the wire that can be changed using gates. The second one represents the surface charge that is typically present at the boundaries of these semiconducting wires \cite{Olsson:PRL96}. This surface charge cannot be removed using gates and it thus gives an intrinsic contribution to the electrostatic potential. In this work we choose $\rho_{\rm{surf}}=5\cdot 10^{-3}\left(\frac{e}{nm^3}\right)$, similarly to previous theoretical works \cite{Winkler:PRB19, Escribano:PRB19, Woods:arxiv19}, and in agreement with experimental evidence \cite{Thelander:Nano10, Reiner:arxiv19}.
This value however does not play any fundamental role but simply results in a small particular contribution to the intrinsic doping of the wire and its SO coupling \cite{Escribano:PRB19}. The numerical methods used to solve Eq. \eqref{Eq:Poisson} are described in App. \ref{App:D}.

We note that the solution of the coupled Schr\"odinger-Poisson equation typically possess a rather demanding numerical problem, as explained in App. \ref{App:D}. However, it can be simplified by relying on the Thomas-Fermi approximation, as shown in previous works for semiconducting nanowires \cite{Mikkelsen:PRX18, Wojcik:PRB18, Escribano:PRB19}. This approximation allows to decouple both equations by assuming that the charge density of the wire is indeed very similar to that of a 3D free electron gas,
\begin{eqnarray}
\rho_{\rm{mobile}}(\vec{r})\simeq\rho_{\rm{mobile}}^{(\rm{TF})}(\vec{r})=  \nonumber \\
-\frac{e}{3\pi^2}\left[\frac{2m^*|e\phi(\vec{r})+E_F|f(-(e\phi(\vec{r})+E_F))}{\hbar^2} \right]^{\frac{3}{2}}
\label{Eq:Charge_density_TF}
\end{eqnarray} 
where $E_F$ is the Fermi energy of the wire and $f(E)$ is the Fermi-Dirac distribution for a given temperature. In the next section we compare the results provided by both approaches in some specific cases, showing that the Thomas-Fermi approximation predicts roughly the same Rashba coupling as Schr\"odinger-Poisson.


\begin{figure}
\includegraphics[scale=0.28]{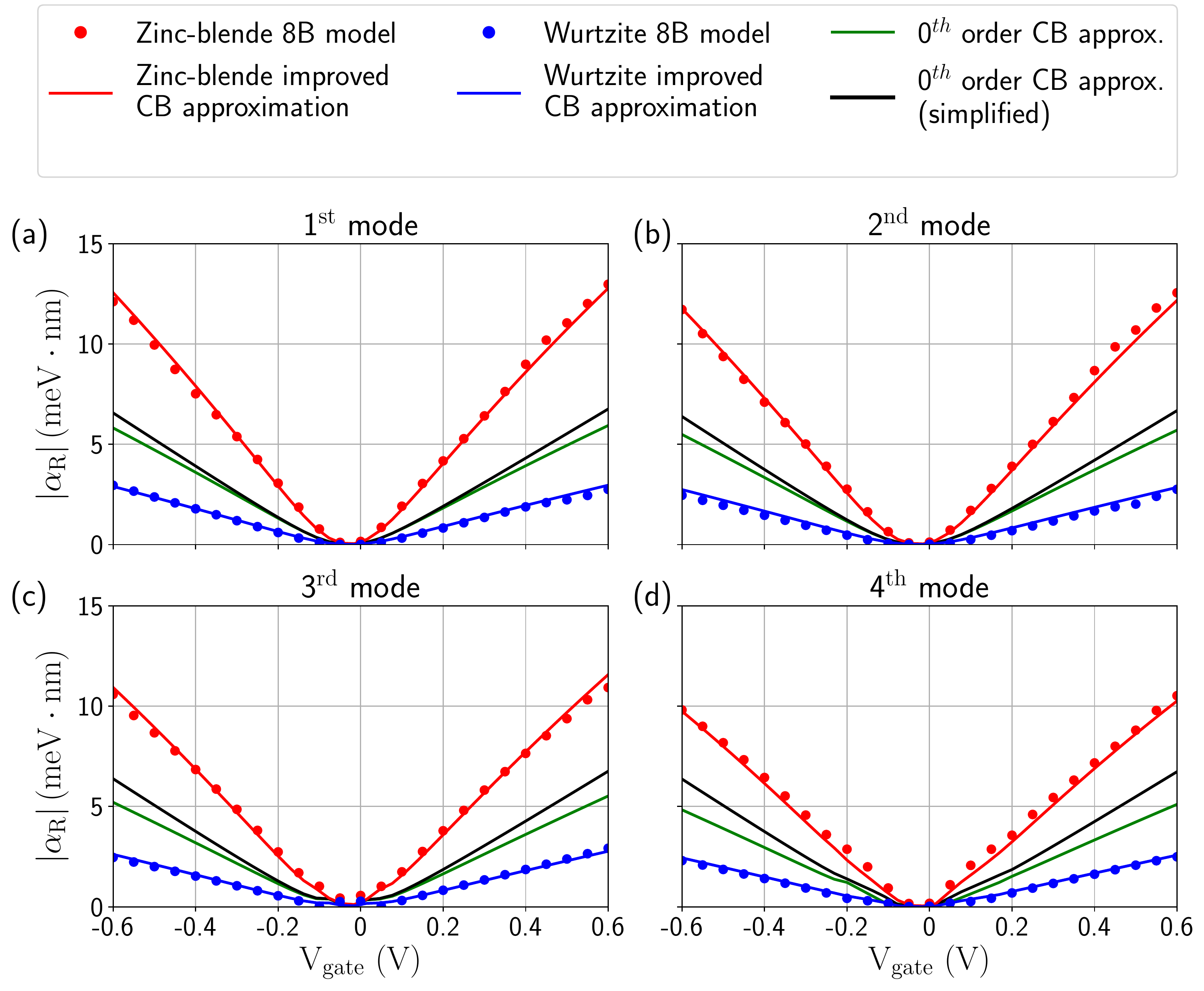}
\caption{Rashba SO coupling modulus versus gate voltage for an InAs nanowire in an electrostatic environment like the one of Fig. \ref{Fig1}(a). Different transverse subbands within the first conduction band of the wire are considered (a-d). Dots correspond to $\alpha_R$ obtained from the eight-band (8B) k$\cdot$p Kane model of App. \ref{App:A} using Eq. \eqref{E_fit_kp}. Solid lines correspond to $\alpha_R$ obtained from the effective conduction band (CB) approximation discussed in this work. In red (blue) the result for a Zinc-blende (Wurtzite) crystal using the improved Eq. \eqref{Eq:alpha_R_improved} with the corresponding parameter $P_{\rm{fit}}$ displayed in Table \ref{Table1}. In green the result using the original Kane parameter $P$. In black the Rashba coupling obtained in the approximation $\Delta_g\gg \left|e\phi(\vec{r})+E\right|$, i.e., using the simplified Eq. \eqref{Eq:alpha_R-oversimplified}. Parameters are: $W_{\rm{wire}}=80$ nm, $W_{\rm{substrate}}=20$ nm, $W_{\rm{layer}}=10$ nm, $\epsilon_{\rm{wire}}=\epsilon_{\rm{InAs}}=15.15$, $\epsilon_{\rm{substrate}}=\epsilon_{\rm{HfO_2}}=25$, $V_{\rm{layer}}=0$V and $\rho_{\rm{surf}}=5\cdot 10^{-3}\left(\frac{e}{nm^3}\right)$. The charge density of the wire $\rho_{\rm{mobile}}$ has been neglected for simplicity. The 8B model parameters are given in Tables \ref{Table2} and \ref{Table3}.}
\label{Fig2}
\end{figure}


\begin{table}[]
\caption{Parameter $P_{\rm{fit}}$ (in meV$\cdot$nm units) to be used in the improved equation for the Rashba SO coupling, Eq. \eqref{Eq:alpha_R_improved}, within the conduction band approximation. This parameter is extracted by fitting Eq. \eqref{Eq:alpha_R_improved} to numerical eight-band model calculations. For comparison, we show the value of the original Kane parameter $P$.}
\begin{tabularx}{0.47\textwidth} {|>{\centering}X||>{\centering}X|>{\centering}X|>{\centering\arraybackslash}X|}
\hline 
      &  Zinc-blende (111)   &  Wurtzite (0001) & Kane $P$       \\ \hline \hline
 InAs &  1252$\pm$12  &  723.0$\pm$0.1  & 919.7 \\ \hline 
 InSb &  1082$\pm$7   &  -   & 940.2            \\ \hline
 GaAs &  1912$\pm$18  &  -   & 1047.5            \\ \hline
 GaSb &  1657$\pm$35  &  -   & 971.3            \\ \hline
\end{tabularx}
\label{Table1}
\end{table} 

\section{Results}
\label{Sec-Results}

We first discuss how to obtain the improved parameter $P_{\rm{fit}}$ of Eq. \eqref{Eq:alpha_R_improved} for Zinc-blende (111) and Wurtzite (0001) III-V compound semiconductors. Specifically, we focus on InAs, GaAs, InSb and GaSb. To do so, as already mentioned, we fit Eq. \eqref{Eq:alpha_R_improved} to calculations based on the 8B model. We consider the geometry depicted in Fig. \ref{Fig1}(a), where an infinite semiconductor nanowire with hexagonal cross section is placed on top of a dielectric substrate, particularly Hf$_2$O. Below the substrate, there is a bottom gate, which allows to tune the chemical potential inside the wire. To increase the gradient of the electrostatic potential, and therefore enhance the SO coupling, the upper facet of the wire is covered by a grounded metallic layer. The motivation for considering this particular environment is that this kind of geometries are typically used in experimental setups for spintronics \cite{Liang:Nano12, Weperen:PRB15, Scherubl:PRB16} and Majorana nanowire devices \cite{Moor:IOP18, Bommer:PRL19}. Since our sole concern in this subsection is to obtain $P_{\rm{fit}}$, we ignore for the moment the electron-electron interactions by fixing $\rho_{\rm{mobile}}=0$.

The modulus \footnote{In our setup, the Rashba SO coupling has only $x$ and $y$ components. The $z$ component is zero due to translational invariance in that direction.} of the Rashba coupling obtained from the 8B model through Eq. \eqref{E_fit_kp} is plotted with dots in Fig. \ref{Fig2} for the case of InAs nanowires. The first four transverse subbands are considered in (a-d), represented versus the back gate voltage. Red (blue) dots correspond to a Zinc-blende (Wurtzite) crystal structure. The corresponding 8B Hamiltonians used for the calculation are provided in App. \ref{App:A}, together with the values of their parameters, displayed in Tables \ref{Table2} and \ref{Table3}.
We can see that the SO coupling exhibits a minimum around $-0.04$V, independently of the transverse mode. This is because, at this gate voltage value, the electric field inside the wire is basically zero. It occurs at $V_{\rm{gate}}\neq0$ due to the surface charge present at the nanowire facets, which introduces a small pinned electrostatic field \cite{Reiner:arxiv19}. As was noticed in previous works \cite{Faria:PRB16}, the SO coupling is larger for the Zinc-blende crystal than for the Wurtzite one, approximately a factor of four.

Now we proceed to fit the previous 8B model results with those obtained with the improved Eq. \eqref{Eq:alpha_R_improved}. We use the values for $\Delta_{g}$ and $\Delta_{\rm{soff}}$ provided in Ref. \onlinecite{Vurgaftman:JAP01} and shown in App. \ref{App:A}. The resulting $P_{\rm{fit}}$ values for Zinc-blende and Wurtzite InAs crystals are collected in Table \ref{Table1}. We represent $|\vec{\alpha}^{(j)}_{R,\rm{improved}}|$ with solid red and blue lines in Fig. \ref{Fig2}. Notice that we can use the same fitting parameter, calculated for the first mode, for all the different transverse modes since $P_{\rm{fit}}$ depends only very slightly on the subband energy. This is not \textit{a priori} obvious since the intra-valence band corrections depend, in principle, on the specific subband. Fortunately, as shown in Fig. \ref{Fig2}, the fit is very good also for higher transverse modes. We further note that the SO coupling decreases with the number of the transverse mode, i.e., it is larger for the lowest energy subband. This can be directly deduced from Eq. \eqref{Eq:alpha_R_improved} since it is inversely proportional to the subband energy $E_T^{(j)}$.

We have performed equivalent calculations for Zinc-blende (111) InSb, GaAs and GaSb nanowires. These can be found in App. \ref{App:E} and the corresponding values of $P_{\rm{fit}}$ in Table \ref{Table1}.

We now compare the previous results for specific crystal structures with the ones that one would obtain directly from the reduced Hamiltonian. In Fig. \ref{Fig2} we show with a solid green line the SO coupling obtained with the zeroth-order Eq. \eqref{Eq:alpha_R} (projecting over $|\Psi_T^{(j)}\rangle$ and with $E$ replaced by $E_T^{(j)}$).
We observe that $\alpha_R^{(0)}$ underestimates the SO coupling for Zinc-blende crystals and overestimates it for Wurtzite ones, approximately by a 50$\%$. This means that the effects of the specific crystal structure are lost in the zeroth-order conduction band approximation.

Finally, the black solid lines in Fig. \ref{Fig2} correspond to simulations using the simplified zeroth-order Rashba coupling, Eq. \eqref{Eq:alpha_R-oversimplified}, frequently used in the literature. We observe that this equation provides not only  different quantitative results, but also qualitative ones, since it cannot capture the variation of $\alpha_R$ with different transverse modes when $E_T^{(j)}$ is not negligible with respect to $\Delta_g$, which is the common experimental situation. Hence, we conclude that both zeroth-order equations do not describe accurately the Rashba SO coupling in type III-V semiconductors.

To carry out the comparison between the different methods described above we needed to consider a specific electrostatic environment. However, the value of $P_{\rm{fit}}$ should be independent of it for the applicability of Eq. \eqref{Eq:alpha_R_improved} in arbitrary conditions. We have performed equivalent fittings for very different environments, see App. \ref{App:F}, finding that $P_{\rm{fit}}$ varies only within a $2\%$ with respect to the value presented in Table \ref{Table1}. This means that $P_{\rm{fit}}$ is reasonably independent of $\phi(\vec{r})$ for typical experimental conditions. The reason is that the main dependence of the SO coupling with the electrostatic potential comes from the gradient of $\phi(\vec{r})$. The prefactor within brackets in Eq. \eqref{Eq:alpha_R_improved} (or in Eq. \eqref{Eq:alpha_R} for that matter) depends on $\phi(\vec{r})$ through the quantity $E+e\phi(\vec{r})$. Ignoring terms proportional to $(\vec{\nabla \phi})^2$, $(\vec{\nabla \phi})^3$, etc., higher order terms in the series expansion of the conduction band approximation satisfy a similar dependence with $\phi(\vec{r})$ but with more complicated prefactors, see for instance an approximation to the first order correction in App. \ref{App:B}, Eq. \eqref{Eq:Rashbaorder1}. The key observation is that all these prefactors depend on the quantity $E+e\phi(\vec{r})$. On the other hand, the energy is the solution to the eigenvalue problem of the conduction band Hamiltonian, $E\sim E_{\rm{K}}-e\langle\phi(\vec{r})\rangle+ E_{\rm{SO}}$, where $E_{\rm{K}}$ stands for the kinetic energy and $E_{\rm{SO}}$ for the Rashba one. Thus the electrostatic potential gets roughly canceled in all those prefactors. Since $P_{\rm{fit}}$ is an approximation to the sum of these series terms, its dependence with $\phi(\vec{r})$ should be small as well.

\begin{figure}
\includegraphics[scale=0.28]{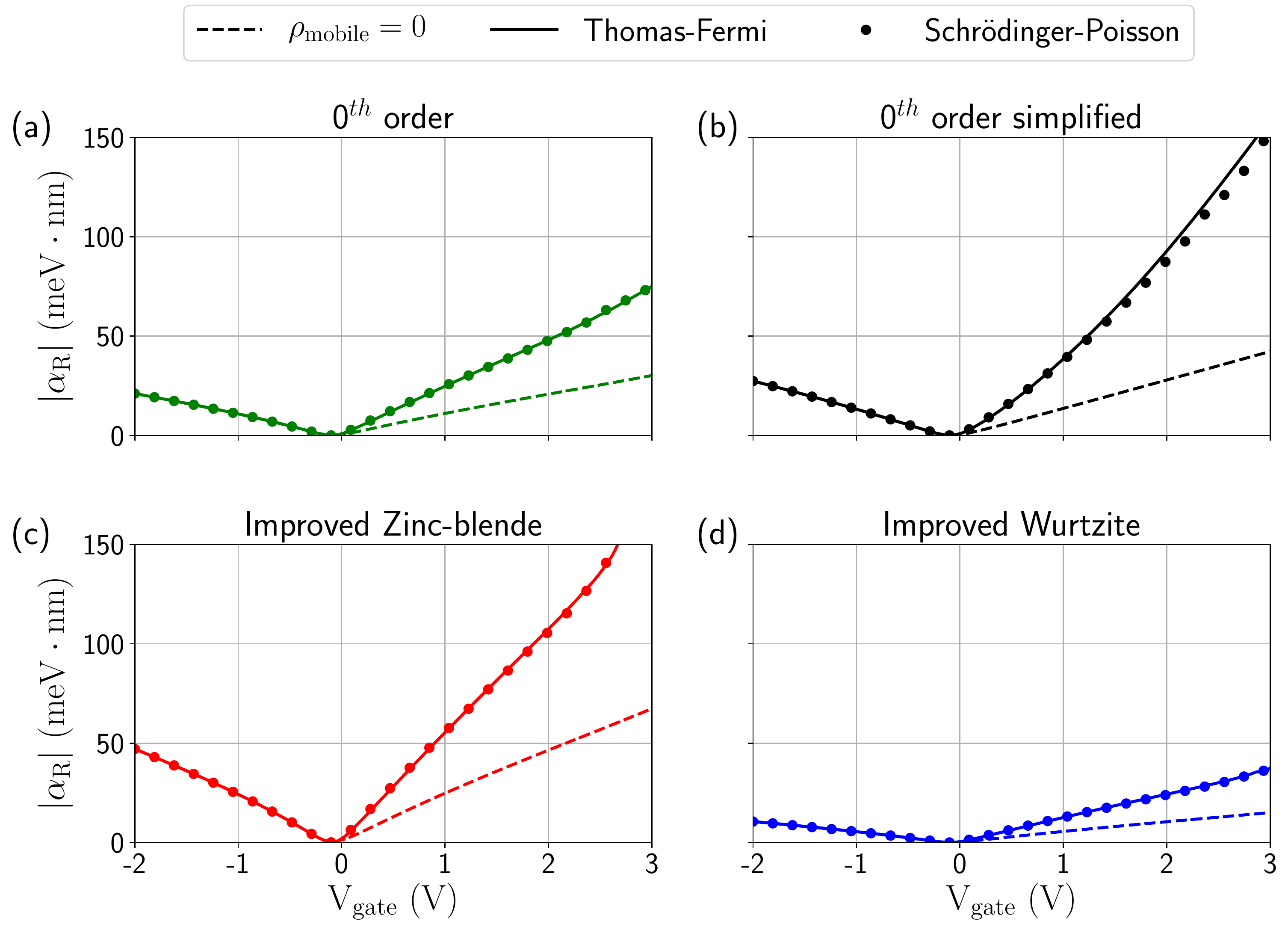}
\caption{Rashba coupling modulus versus gate voltage for the lowest energy subband within the conduction band approximation. (a) In green the results obtained using the improved  Eq. \eqref{Eq:alpha_R_improved} but with the original Kane parameter $P$. (b) In black, using the simplified Eq. \eqref{Eq:alpha_R-oversimplified}. (c,d) In red and blue the results of the improved equation using the corresponding $P_{\rm{fit}}$ parameter displayed in Table \ref{Table1} for a Zinc-blende and a Wurtzite crystal, respectively. Dots are used for full Schr\"odinger-Poisson simulations, solid lines for the Thomas-Fermi approximation and dotted lines for $\rho_{mobile}=0$, i.e., ignoring the mobile charge density. Same parameters as the ones of Fig. \ref{Fig2} (except $\rho_{\rm{mobile}}$). Temperature $T=10$ mK.}
\label{Fig3}
\end{figure}

\subsection{Schr\"odinger-Poisson vs Thomas-Fermi}
The previous simulations were performed ignoring the charge density of the wire, $\rho_{\rm{mobile}}$. Including it requires a self-consistent solution of the coupled Schr\"odinger-Poisson equations. Especially for the 8B model, this is a difficult task due to the large number of bands involved \cite{Ehrhardt:14}.
As  explained in Sec. \ref{Sec-Electrostatic_environment}, the Thomas-Fermi approximation helps to reduce the computational cost of these simulations.

In order to test the validity of the Thomas-Fermi approximation for the determination of of the SO coupling and whether the electron-electron interaction plays a relevant role, we have performed the same simulations as before in the conduction band approximation but including $\rho_{\rm{mobile}}$ \footnote{We ignore here for simplicity the contribution to $\rho_{\rm{mobile}}$ coming from the valence band. Its inclusion would only affect the value of $\alpha_R$ at very negative gate voltages.}. In Fig. \ref{Fig3} we show the SO coupling as a function of gate voltage in the absence of mobile charge (dashed line), including it in the Thomas-Fermi approximation (solid line) and using the full Schr\"odinger-Poisson approach (dots). We find that  Thomas-Fermi provides an excellent approximation that matches the Schr\"odinger-Poisson results for Zinc-blende, Wurtzite, as well as simplified structures. Moreover, we see that taking into account the charge density of the wire is essential to describe the behavior of $\alpha_R$ for large positive $V_{\rm{gate}}$ values (large doping). This is due to the inhomogeneous electrostatic potential profile created by the charge, which contributes to the structural inversion asymmetry that in turn leads to the Rashba coupling. Once again, we note that the quantitative results predicted by the simplified equation (black curves) deviate considerably from the ones predicted for specific crystal structures (red or blue curves).

\subsection{Confinement effects}

As explained in App. \ref{App:B}, the terms that correct the zeroth-order Rashba coefficient of Eq. \eqref{Eq:alpha_R} in the conduction band approximation are proportional to $\sim\hbar^2\vec{k}^2\gamma_i/(2 m_e)$, where $m_e$ is the electron mass and $\gamma_{i}$ are the intra-valence band couplings. Since these terms are proportional to the transverse momenta, $k_x,k_y$, they produce SO corrections coming from each transverse subband in finite-width nanowires.

\begin{figure}
\includegraphics[scale=0.28]{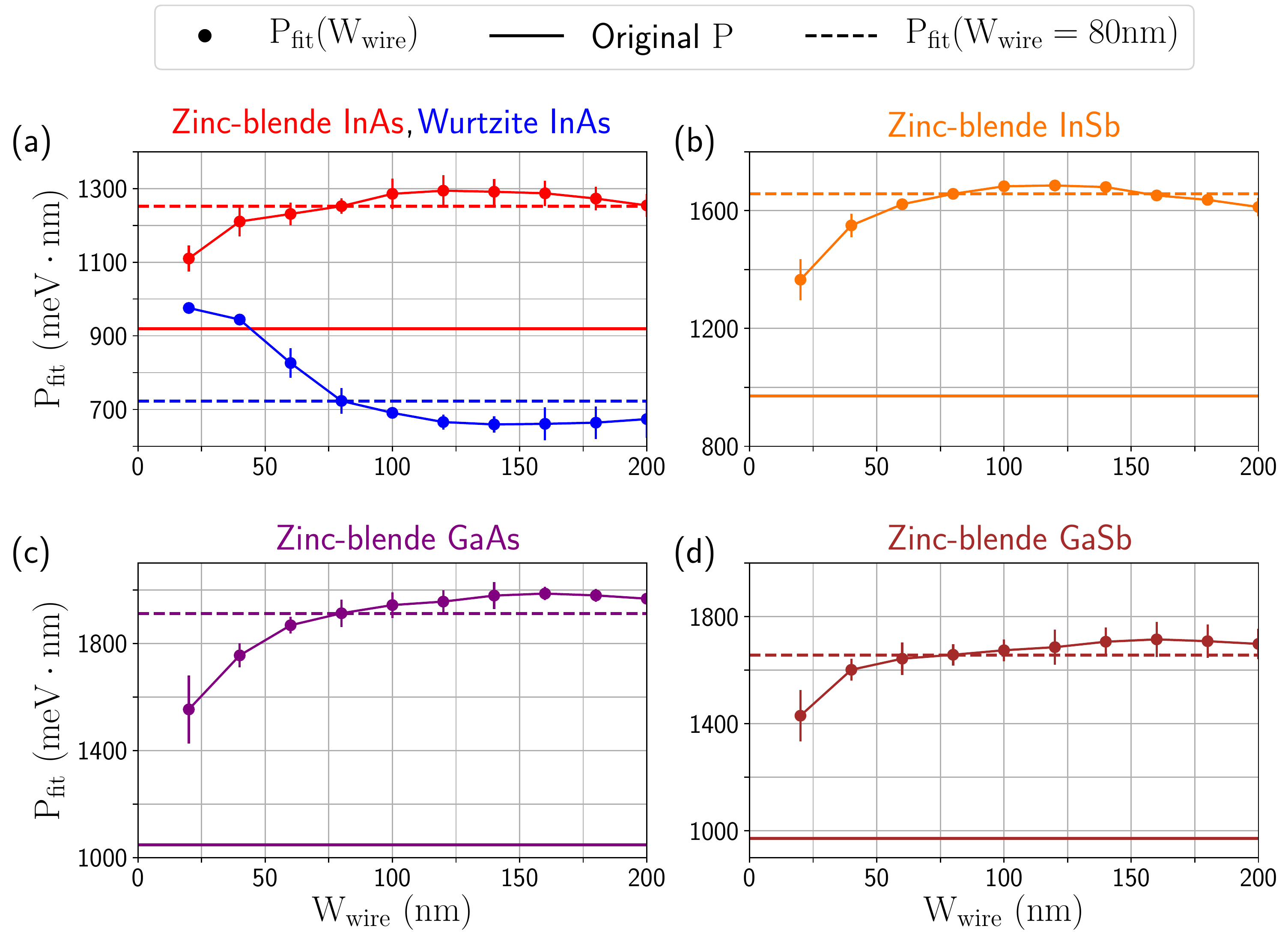}
\caption{Fitting parameter $P_{\rm{fit}}$ of Eq. \eqref{Eq:alpha_R_improved} as a function of the nanowire's width $W_{\rm{wire}}$ (curve with dots) for different semiconductors and crystal structures. This parameter is extracted by fitting Eq. \eqref{Eq:alpha_R_improved} to 8B model results. The electrostatic environment corresponds to the one of Fig. \ref{Fig1}(a) with parameters as in Fig. \ref{Fig2}, but changing the dielectric permittivity of the wires to their corresponding ones, particularly $\epsilon_{\rm{InAs}}=15.15$, $\epsilon_{\rm{InSb}}=16.8$, $\epsilon_{\rm{GaAs}}=12.9$ and $\epsilon_{\rm{GaSb}}=15.7$, extracted from Ref. \onlinecite{Levinshtein:00}. For comparison, with a dashed line we show $P_{\rm{fit}}$ for a specific diameter, $W_{\rm{wire}}=80$ nm, which is the tabulated value given in Table \ref{Table1}, and with a solid line we show the original Kane parameter $P$.}
\label{Fig4}
\end{figure}

To quantify this effect in Fig. \ref{Fig4} we consider wires of different width $W_{\rm{wire}}$ in an electrostatic environment like the one of Fig. \ref{Fig1}(a). By using the improved Eq. \eqref{Eq:alpha_R_improved} and by fitting it to 8B model calculations, we plot the value of the resulting $P_{\rm{fit}}$ parameter as a function of $W_{\rm{wire}}$. This is done for Zinc-blende/Wurtzite InAs, and Zinc-blende InSb, GaAs and GaSb. We observe that the fitting parameter for Zinc-blende structures increases for small diameters, has a maximum around $100-150$ nm (depending on the crystal) and then slowly decreases as $W_{\rm{wire}}\rightarrow\infty$, approaching the original Kane parameter $P$ depicted with a solid line (the opposite happens for Wurtzite structures, with a minimum instead of a maximum). The value of $P_{\rm{fit}}$ is nevertheless pretty constant for a wide range of wire's widths, which correspond to the common experimental values. This justifies our approximation of using a \emph{simple} width-independent $P_{\rm{fit}}$ parameter in our ansatz for an improved SO coupling equation. In particular, we consider $P_{\rm{fit}}(W_{\rm{wire}})=80$ nm, represented with a dashed line in Fig. \ref{Fig4}, which is a good approximation of $P_{\rm{fit}}(W_{\rm{wire}})$ for diameters between $\sim50$ nm and $\sim200$ nm.

In conclusion, in finite-width nanowires with non-negligible transverse momenta $k_x,k_y\sim 1/W_{\rm{wire}}$, the higher order terms neglected in the zeroth-order conduction band approximation turn out to be important to predict the correct SO coupling. The reason is that different transverse subbands acquire modified gaps proportional to $\gamma_i$ at the $\Gamma$ point. Moreover, the dependence with $\gamma_i$ produces dramatic differences between distinct crystal structures (Zinc-blende vs Wurtzite). These confinement effects disappear as $W_{\rm{fit}}\rightarrow\infty$, rendering the zeroth-order approximation correct for bulk crystals.

\section{Comparison to experiments}
\label{Sec-Applications}
In order to test the validity of our approach to predict the SO coupling in realistic situations, we compare the results that it provides with the ones obtained in several recent experiments. During the last decade, there has been an increasing interest in measuring the SO coupling in semiconductor nanowires due to their potential applications as spin-FETs \cite{Hansen:PRB05, Dhara:PRB09, Liang:Nano12, Scherubl:PRB16, Takase:SciRep17,Iorio:NanoLett19} or Majorana qubit devices \cite{Moor:IOP18, Bommer:PRL19}. In most cases, these nanowires were made of InAs due to its large semiconductor band gap and Rashba coupling, although some of them used InSb nanowires \cite{Weperen:PRB15, Takase:IOP19} and other mixed heterostructures \cite{Nitta:PRL97,Zumbuhl:PRL02, Studenikin:PRB03, Schapers:PRB04, Koo:Science09, Kallaher:PRB10, Nadj-Perge:PRL12, Zellekens:arxiv19} involving type III-V compound semiconductors. For the comparison, we focus on the works done by Dhara \textit{et al.} \cite{Dhara:PRB09}, Liang \textit{et al.} \cite{Liang:Nano12}, Takase \textit{et al.} \cite{Takase:SciRep17}, and Scher\"ubl \textit{et al.} \cite{Scherubl:PRB16}, carried out on Zinc-blende InAs nanowires; and by Takase \textit{et al.} \cite{Takase:IOP19} on Zinc-blende InSb nanowires. We choose these experiments because they measure a representative number of SO coupling points versus a wide range of gate potentials. 

In all of these experiments, the SO coupling is determined in an indirect way from magnetotransport measurements \cite{Chakravarty:PhyRep86, Beenaker:PRB88, Kurdak:PRB92}, which permit to access relevant length scales that affect the electron coherence.
In particular, the SO coupling can be extracted from the spin-relaxation length $l_{\rm{SO}}$ as $\alpha_{\rm{eff}}=\frac{\hbar^2}{2m_{\rm{eff}}l_{\rm{SO}}}$. Hence, to extract the SO coupling, it is necessary to know the electron's effective mass. In all the experiments that we analyze here, the authors use the same values, $m_{\mathrm{eff}}=0.023m_e$ for InAs and $m_{\mathrm{eff}}=0.014m_e$ for InSb. These are precisely the values one gets from the zeroth-order simplified Eq. \eqref{Eq:m_eff-oversimplified} with the original Kane $P$ parameter. One could argue that, the same way we need to improve the zeroth-order equation for the SO coupling, as we demonstrate in this work, one would also need to correct the effective mass $m^{(0)}$ to match the 8B model results. In App. \ref{App:G} we show that this is indeed not necessary. The zeroth-order is already a very good approximation to the results provided by the 8B model Hamiltonians. This is further true as a function of the wire's width, as we analyze in Fig. \ref{FigB}(b) of App. \ref{App:B}. Therefore, the wide spread use of the previous values for the effective masses, as well as the equivalent ones for other semiconductor compounds, is completely justified.

\begin{figure}
\includegraphics[scale=0.265]{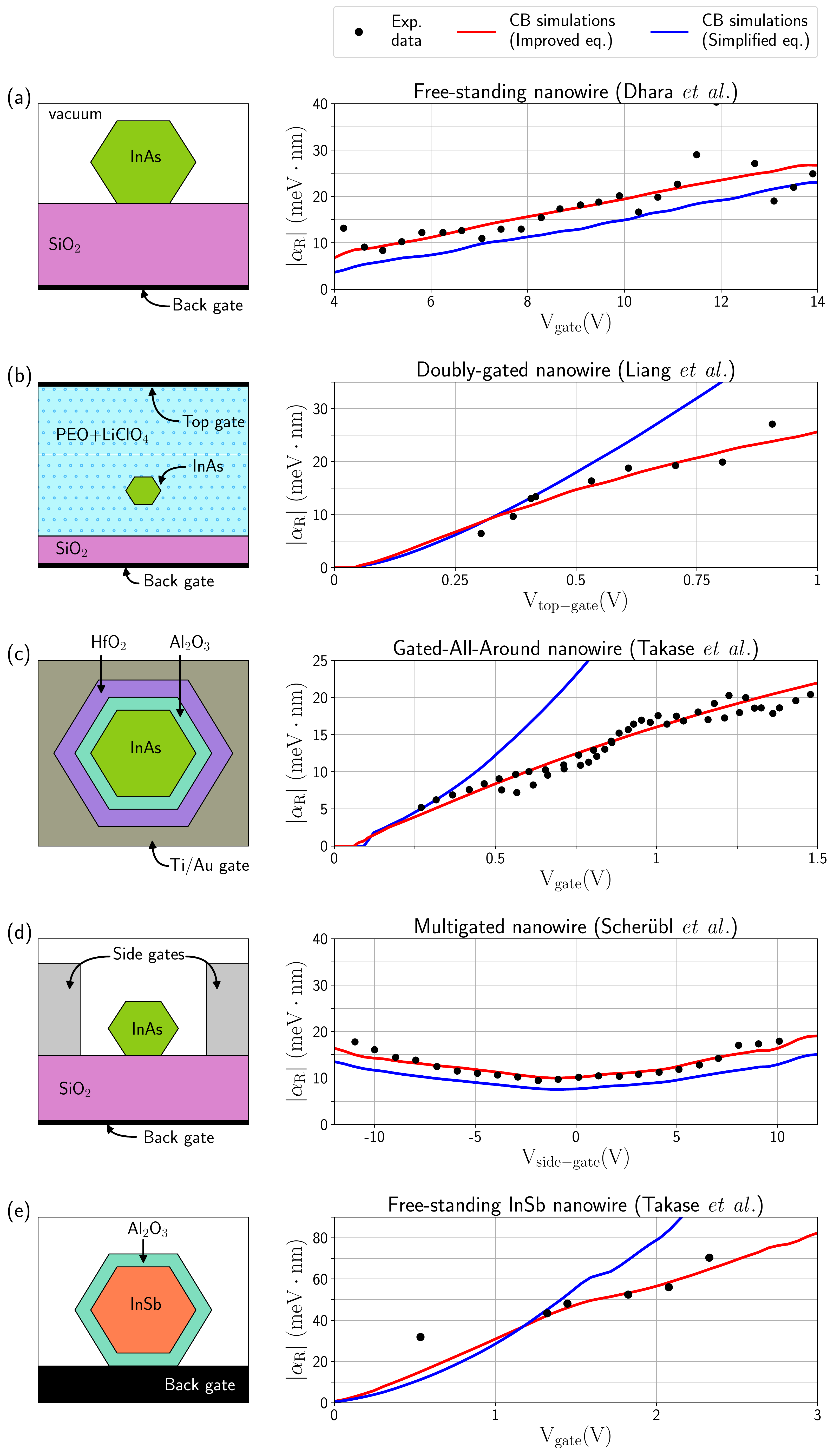}
\caption{Electrostatic environment modeling of some experimental setups (left), and corresponding effective Rashba couplings (right) obtained with magnetoconductance measurements (dots) and with conduction band (CB) numerical simulations (solid lines). In red we present results using the improved Eq. \eqref{Eq:alpha_R_improved} for Zinc-blende crystals, while in blue using the simplified Eq. \eqref{Eq:alpha_R-oversimplified}. The   shown experimental data corresponds to (a) Dhara \textit{et al.} \cite{Dhara:PRB09}, (b) Liang \textit{et al.} \cite{Liang:Nano12}, (c) Takase \textit{et al.} \cite{Takase:SciRep17}, (d) Scher\"ubl \textit{et al.} \cite{Scherubl:PRB16}, and (e) Takase \textit{et al.} \cite{Takase:IOP19}. The charge density of the wire is taken into account in the Thomas-Fermi approximation and the surface charge is $\rho_{\rm{surf}}=5\cdot 10^{-3}\left(\frac{e}{nm^3}\right)$. The geometrical parameters used in the simulations can be found in the main text. The dielectric constants, extracted from Refs. \onlinecite{Levinshtein:00, Robertson:EPJAP04,DAS:AIP15}, are $\epsilon_{\rm{InAs}}=15.15$, $\epsilon_{\rm{InSb}}=16.8$, $\epsilon_{\rm{SiO_2}}=3.9$, $\epsilon_{\rm{HfO_2}}=25$, $\epsilon_{\rm{Al_2O_3}}=9$ and $\epsilon_{\rm{PEO}}=10^4$. In agreement with the experiments, in our simulations we fix the temperature to $T=1.7$K in (a), $T=4$K and the back gate to 0 in (b), $T=1.5$K in (c), $T=4.2$K and the back gate to 15V in (d), and $T=1.7$K in (e). }
\label{Fig5}
\end{figure}

On the other hand, since this kind of measurements involves the collective transport of electrons around the Fermi energy, the SO coupling extracted from $l_{\rm{SO}}$ does not correspond to one particular subband but it is instead a weighted sum of all the subbands that contribute to the current. To compute numerically this averaged Rashba coupling in the conduction band approximation, we take the expected value of the SO coefficient
\begin{equation}
\vec{\alpha}_{\rm{eff}}^{(\rm{EV})}=\frac{\sum_j \left<\vec{\alpha}_{R}^{(j)}(\vec{r})\right> n^{(j)}}{\sum_j n^{(j)}},
\end{equation}
where $n^{(j)}$ is the occupation of transverse subband $j$ (see App. \ref{App:C} for further details). In Fig. \ref{Fig5} we compare the experimental data (dots) with the numerical results obtained with the improved Eq. \eqref{Eq:alpha_R_improved} and the $P_{\rm{fit}}$ values of Table \ref{Table1} (red curves). The electrostatic potential is calculated using the Thomas-Fermi approximation. For completeness, we also show the results provided by the simplified Eq. \eqref{Eq:alpha_R-oversimplified} (blue curves).

Figure \ref{Fig5}(a) refers to the experiment performed by Dhara \textit{et al.} \cite{Dhara:PRB09}. This is one of the first works that used magnetotransport measurements to determine the Rashba coupling in InAs nanowires. The device is quite simple (see sketch on the left): an 80 nm wide InAs nanowire is placed on top of a SiO$_2$ substrate and 300 nm below the substrate there is a bottom gate that is used to tune the electrostatic potential inside the wire.
The large $V_{\rm{gate}}$ range explored in this experiment with a relatively small variation of the SO coupling, see right panel, is due to the rather large thickness of the substrate. 
Both theoretical curves predict the same qualitative behavior but, although they are quantitatively similar, our approach gives a somewhat better agreement with the experimental values. For this particular setup, the simplified equation works reasonably well because the electrostatic potential created by the gate is small. Thus the condition $\Delta_g\gg \left|e\phi(\vec{r})+E\right|$ is almost fulfilled.

Figure \ref{Fig5}(b) refers to the experiment carried out by Liang \textit{et al.} \cite{Liang:Nano12}. They prove that it is possible to enhance the Rashba coupling by using an appropriate electrostatic environment. To do so, a 40 nm wide InAs nanowire is suspended inside the (ionic) dielectric PEO+LiClO$_4$. At 50 nm below the wire there is a SiO$_2$ substrate of thickness 250 nm sitting on a grounded gate. On top of the device, 500 nm above the wire, there is another gate in contact with the dielectric. The PEO+LiClO$_4$ dielectric is characterized by a large permittivity, which allows to subject the wire boundaries to almost the same potential as it is applied to the top gate. It is thus possible to significantly increase the wire's doping with a small gate voltage. The origin of the SO coupling in this setup is the inhomogeneous distribution of the charge along the radius of the wire, which is sometimes called pair SO coupling in the literature \cite{Gindikin:EPJ20}.
In the right panel we can see that the red curve is in good agreement with the experimental data, whereas the blue one deviates, especially at large $V_{\rm{gate}}$, due to the large electrostatic potentials involved.

Figure \ref{Fig5}(c) deals with the experiment done by Takase \textit{et al.} \cite{Takase:SciRep17}. This work follows the same spirit than the previous one, but they look for a long-lived device that could be used as a spin-FET for spintronic applications. To this end, they fabricate a gated-all-around (GAA) device in which a 100 nm wide nanowire is covered with a 2 nm thick Al$_2$O$_3$ dielectric, a 4 nm thick HfO$_2$ dielectric and a potential gate (see sketch on the left). Due to the small distance between the gate and the wire, the potential applied to the gate and at the boundary of the wire is almost the same. As in the previous example, the theoretical prediction resulting from our improved equation exhibits an excellent agreement with the experimental data, while the simplified one largely deviates from it. In this case, and motivated by their transport measurements, we have chosen the Fermi level at $E_F=-100$ meV, as if the wire was initially doped with holes.

Figure \ref{Fig5}(d) shows the last case with InAs. It was performed by Scher\"ubl \textit{et al.} \cite{Scherubl:PRB16} to prove that it is possible to tune the Rashba coupling without changing the electron occupation inside the nanowire. To that end, they use a bottom gate together with two side ones.
As shown in the sketch, a 77 nm wide nanowire is placed over a SiO$_2$ substrate, a bottom gate is 1 $\mu$m below the substrate and the two side gates are placed at 70 nm from the corners of the wire. The main origin of the strong SO coupling in this device is the structural inversion symmetry created by the difference between the gate potentials applied to the side gates. 
In the right panel, the SO coupling is plotted versus one side gate, while the other side gate is changed accordingly to keep the total charge inside the wire constant (see Ref. \onlinecite{Scherubl:PRB16} for further information). We find here that both theoretical methods give a good match, although our improved equation produces a better agreement with the data. Actually, in order to explain the discrepancy between the prediction of the simplified equation and the experimental data, in Ref. \onlinecite{Scherubl:PRB16} the authors add \textit{ad hoc} a built-in intrinsic SO coupling of $\sim5$ meV$\cdot$nm, which is not necessary using our effective equation.

Finally, Fig. \ref{Fig5}(e) refers to the experiment realized by Takase \textit{et al.} \cite{Takase:IOP19} in InSb nanowires. This work proves that the SO coupling of InSb nanowires is much larger than that of InAs ones. Their device consists of a 182 nm wide wire covered by a 6 nm thick Al$_2$O$_3$ dielectric, placed directly over a metallic gate. The small thickness of the dielectric allows to tune (almost) perfectly the wire. 
In this last case we also find a good agreement between the red curve and the experimental measurements, while the simplified equation fails to predict the proper behavior. Deviations at small $V_{\rm{gate}}$ may arise due to the small length of the wire in this experiment (500 nm). This may cause that the leads used for the transport measurements have an impact on the electrostatic potential profile at the wire edges \cite{Escribano:BJN18}, changing unintentionally the precise SO coupling value.

\section{Summary and Conclusions}
\label{Sec-Conclusions}

Multiband k$\cdot$p effective models are successfully used to indirectly extract the SO coupling of semiconductors from their band structure, but they can be computationally demanding in low-dimensional heterostructures of current interest, subject to arbitrary electrostatic environments. In this work, we perform a single band approximation and introduce a heuristic analytical expression that accurately describes the conduction band Rashba SO coupling of III-V semiconductor nanowires, Eq. \eqref{Eq:alpha_R_improved}. This equation takes into account not only the dependence of $\vec{\alpha}_R$ with the spatially-dependent electrostatic potential, which accounts for possible structural inversion asymmetries, but also with the transverse subband energy. It further depends on two semiconductor parameters, the band gap and the split-off gap between valence bands. Additionally, it approximately takes into account the crystal structure of the compound semiconductor, partially lost in the usual the zeroth-order conduction band approximation, through one improved effective parameter that we call $P_{\rm{fit}}$. This parameter substitutes the original Kane parameter $P$ in order to account, in a simple and manageable way, for sizable SO coupling contributions originating from transverse subbands in finite-width nanowires. We compute this parameter by fitting the SO coupling given by Eq. \eqref{Eq:alpha_R_improved} to results provided by realistic 8B k$\cdot$p calculations. The results for Zinc-blende (111) InAs, InSb, GaAs and GaSb nanowires, and for Wurtzite (0001) InAs nanowires are collected in Table \ref{Table1}. Using these numbers, we find that the magnitude of the SO coupling of nanowires based on Zinc-blende InAs is roughly four times stronger than on Wurtzite InAs, whereas the SO coupling of InSb nanowires is three times larger than that of InAs ones. 

We compare the results provided by Eq. \eqref{Eq:alpha_R_improved} with those obtained with other approximations and with exact 8B model calculations. Our improved equation works well regardless of the transverse mode, the particular electrostatic environment surrounding the wire and its chemical composition, considering its specific $P_{\rm{fit}}$ value given in Table \ref{Table1}. It also works well for wire widths ranging form $\sim 50$ nm to $\sim 200$ nm, which are the typical experimental values. In particular, for this range, we find that the usual zeroth-order approximation for the SO coupling underestimates its magnitude for Zinc-blende nanowires and overestimates it for Wurtzite ones, by as much as $\sim 50\%$.

As a final proof of the validity of our approach, we simulate the experimental conditions of five magnetoconductance experiments on InAs and InSb nanowires realized in recent years. We find that the  Rashba SO coupling resulting from Eq. \eqref{Eq:alpha_R_improved} is in excellent agreement with the experimental data over wide ranges of external gate potentials.

We believe that our work may be useful for reducing the computational cost of accurate Rashba SO coupling computations in realistic low-dimensional semiconductors of current interest, where confinement effects turn out to be essential for the correct prediction of $\alpha_R$. It may also help to understand and design better devices where the Rashba SO coupling is key, e.g by tuning appropriately the wire's diameter and in this way maximizing the SO interaction.

The dataset and scripts required to plot the figures of this manuscript can be found in the Zenodo repository \cite{Zenodo}.

\begin{acknowledgments}
The authors would like to thank Eduardo J. H. Lee for valuable discussions. Research supported by the Spanish Ministry of Economy and Competitiveness through Grants FIS2016-80434-P, BES-2017-080374 and FIS2017-84860-R (AEI/FEDER, EU), the European Union's Horizon 2020 research and innovation programme under the FETOPEN grant agreement No 828948, the Ram\'on y Cajal programme RYC-2011-09345 and the Mar\'ia de Maeztu Programme for Units of Excellence in R\&D (CEX2018-000805-M).
\end{acknowledgments}


\bibliography{Rashba}

\onecolumngrid
\appendix

\section{8-band k$\cdot$p Kane model}
\label{App:A}


\subsection{Derivation of the model}
We provide here an introduction to the multiband k$\cdot$p theory extracted from several references \cite{Winkler:03, Willatzen:09, Bastos:IOP16}. This theory can be derived starting from the single-electron Hamiltonian
\begin{eqnarray}
H=\frac{\vec{p}^2}{2m_e}+\frac{\hbar}{4m_e^2c^2}\left[\vec{\nabla}\phi(\vec{r}) \times \vec{p}\right]\cdot\vec{\sigma}-e\phi(\vec{r}) \rightarrow H\Psi=E\Psi,
\label{Eq:Hamiltonian_FreeElectron}
\end{eqnarray} 
where $\vec{p}=-i\hbar\vec{\nabla}$ is the momentum operator, $m_e$ the (bare) electron mass, $c$ the speed of light and $\vec{\sigma}=(\sigma_x,\sigma_y,\sigma_z)$ the vector of Pauli matrices for the spin degree of freedom. The first term of this Hamiltonian corresponds to the kinetic energy of the electron, the second one takes into account relativistic SO effects, and the last one corresponds to the effective electrostatic potential energy experienced by the electron inside some material. If the crystal is translational invariant, this potential can be described using a periodic function. As a consequence, the electron wave function satisfies the Bloch's theorem, i.e.,
\begin{equation}
 \Psi_{n,k}= u_{n,k}(\vec{r})e^{i\vec{k}\cdot\vec{r}},
\end{equation} 
where $\vec{k}$ is the wavevector restricted to the first Brillouin zone (and $\hbar\vec{k}$ is the so-called crystal momentum for a periodic system), $n$ the quantum number that labels the different possible energy bands, and $u_{n,k}(\vec{r})$ is the so-called envelope function that encodes the periodic part of the wavefunction. Hence, the electron wave function satisfies the condition $
\vec{p}\Psi_{n,k}=e^{i\vec{k}\cdot\vec{r}}(\hbar \vec{k}+\vec{p})u_{n,k}(\vec{r})$, what allows to write a Hamiltonian $H_{\rm{kp}}$ for the envelope function
\begin{eqnarray}
H\Psi=E\Psi \rightarrow H_{\rm{kp}} u_{n,k}=E_{n,k} u_{n,k} \rightarrow  \nonumber \\
H_{\rm{kp}}=
\frac{\vec{p}^2}{2m_e}-e\phi(\vec{r})+\frac{\hbar}{4m_e^2c^2}\left[\vec{\nabla}\phi(\vec{r})\times \vec{p}\right]\cdot\vec{\sigma}
+ \hbar\frac{\vec{k}\cdot\vec{p}}{m_e}+\frac{\hbar^2\vec{k}^2}{2m_e}+\frac{\hbar^2}{4m_e^2c^2}\left[\vec{\nabla}\phi(\vec{r})\times \vec{k}\right]\cdot\vec{\sigma}.
\end{eqnarray}
This Hamiltonian is known as the k$\cdot$p Hamiltonian (because it includes a $\vec{k}\cdot\vec{p}$ term). It describes the motion of an electron inside a periodic crystal.

In general, the k$\cdot$p Hamiltonian has no analytical solution, so it is usually solved perturbatively. The most common way to do so is to expand the Hamiltonian around a point $\vec{k}_0$ in reciprocal space, whose solution is known, and then use L\"owdin perturbation theory \cite{Lowdin:ChemPhys51} to perform the expansion over a reduced basis set. Within this technique, the states are separated into two classes A and B. Class A includes the truncated basis set elements that describe the main aspects of the crystal. In principle, this basis set could include all the orbitals on each atom of the unit cell, but this would not help to decrease the complexity of the problem. Because of that, the less influential states not included in  A are collected in class B. The couplings between class B and A states are the ones treated perturbatively.

In this representation, the envelope function (in Dirac notation) is written as a superposition of both kind of states
\begin{equation}
\left|n,k\right>=\sum_\alpha c_{\alpha,n}(k)\left|\alpha\right>+\sum_\beta c_{\beta,n}(k)\left|\beta\right>,
\end{equation} 
where $\left|\alpha\right>$ and $\left|\beta\right>$ are the states in class A and B, respectively. The projection of the Hamiltonian onto the class A basis gives rise to the matrix elements
\begin{eqnarray}
\left<\alpha\right|H_{\rm{kp}}\left|\alpha'\right>=\left[E_\alpha(\vec{k}_0)+\frac{\hbar^2(\vec{k}^2-\vec{k}^2_0)}{2m_e}\right]\delta_{\alpha\alpha'}+\left<\alpha\right|\frac{\hbar}{m_e}(\vec{k}-\vec{k}_0)\cdot \vec{p}\left|\alpha'\right>+\left<\alpha\right|\frac{\hbar}{4m_e^2c^2}\left[\vec{\nabla}\phi(\vec{r})\times \vec{p}\right]\cdot\vec{\sigma}\left|\alpha'\right> \nonumber\\
+\sum_{\beta}\frac{\left<\alpha\right|\frac{\hbar}{m_e}(\vec{k}-\vec{k}_0)\cdot \vec{p} \left|\beta\right>\left<\beta\right|\frac{\hbar}{m_e}(\vec{k}-\vec{k}_0)\cdot \vec{p}\left|\alpha'\right>}{E_\alpha-E_\beta}.
\end{eqnarray}
Unfortunately, these matrix elements cannot be evaluated analytically in general because one would need to know all the dipole terms $\left<\alpha\right|k\cdot p \left|\beta\right>$ of all the transitions among the different states, as well as their corresponding transition energies $E_{\alpha,\beta}$. For this reason, one firstly invokes symmetry arguments (related to the crystal symmetries) to know which matrix elements are forbidden and, secondly, one substitutes the remaining expressions by parameters, whose functional forms can be determined using group theory arguments. These parameters are called Luttinger or Kane parameters, and can be extracted from experimental data or \textit{ab initio} calculations.

In this work, we have applied this technique to compute the electronic band structure of III-V binary compound semiconductors, specifically InAs, InSb, GaAs and GaSb. These are known to have a large Rashba coupling. As the unperturbed Hamiltonian, we choose the one at the $\Gamma$-point ($k_0=0$). This is because this kind of semiconductors are known to have a direct band gap \cite{Vurgaftman:JAP01} at that point, with the minimum of the conduction band taking place there. For our representation basis, we have considered as class A the topmost six electronic states in the valence band (usually referred to as p-like states), given by the heavy hole $\left|HH\right>$, light hole $\left|LH\right>$ and split-off $\left|SO\right>$ bands, and the first two states at the conduction band (usually referred to as s-like states), given by $\left|C\right>$. Note that all these bands are doubly degenerate at the $\Gamma$ point, this is why there are eight in total. For these tetravalence semiconductor materials, it has been shown \cite{Luttinger:PR55, Kane:JPCS57, Faria:PRB16} that these eight bands are enough to describe the main electronic properties and, specifically, the Rashba SO interaction. The relation between these band eigenstates and the (s-type and p-type) atomic orbitals depends on the particular crystal structure of the selected material. In the following subsections, we provide these relations together with the matrix elements of the Hamiltonian for Zinc-blende and Wurtzite crystals, which are the most common crystal structures for these kind of semiconductors.

\subsection{8-band Zinc-blende Hamiltonian}

The 8-band (8B) k$\cdot$p Kane Hamiltonian for Zinc-blende crystals has been derived and described in previous works \cite{Kane:JPCS57, Luo:AIP16}. The basis for Zinc-blende-type semiconductors is given by
\begin{eqnarray}
\left\{ \begin{array}{ll}
\left| C_{\uparrow} \right>=\left| S_{\uparrow} \right> , &
\left| C_{\downarrow} \right>=\left| S_{\downarrow} \right>, \\
\left| LH_{\uparrow} \right>=\frac{i}{\sqrt{6}}\left| (X+iY)_{\downarrow} -2Z_{\uparrow}\right>, &
\left| LH_{\downarrow} \right>=\frac{1}{\sqrt{6}}\left| (X-iY)_{\uparrow} +2Z_{\downarrow}\right>, \\
\left| HH_{\uparrow} \right>=\frac{1}{\sqrt{2}}\left| (X+iY)_{\uparrow} \right>, &
\left| HH_{\downarrow} \right>=\frac{i}{\sqrt{2}}\left| (X-iY)_{\downarrow} \right>, \\
\left| SO_{\uparrow} \right>=\frac{1}{\sqrt{3}}\left| (X+iY)_{\downarrow} +Z_{\uparrow}\right>, &
\left| SO_{\downarrow} \right>=\frac{i}{\sqrt{3}}\left| -(X-iY)_{\uparrow}+Z_{\downarrow} \right>,
\end{array}
   \right.
\end{eqnarray}
where $S$, $X$, $Y$ and $Z$ denote the type of symmetry ($s$-function, or $x$, $y$ or $z$ $p$-function) that the orbital has under the tetrahedral group transformation, and $\left\{\uparrow,\downarrow\right\}$ denotes the spin projection. In this basis, the $z$-direction corresponds to the $(111)$ crystallographic orientation. The Hamiltonian in the $\Psi=(\Psi_{\rm{c},\uparrow},\Psi_{\rm{c},\downarrow},\Psi_{\rm{lh},\uparrow},\Psi_{\rm{hh},\uparrow},\Psi_{\rm{hh},\downarrow},\Psi_{\rm{lh},\downarrow},\Psi_{\rm{soff},\uparrow},\Psi_{\rm{soff},\downarrow})$ basis is given by
\begin{equation}
\label{ZBHamiltonian}
H_{\rm{kp}}=
  \begin{pmatrix}
    T_{\rm{c}} & 0 & \frac{1}{\sqrt{6}}Pk_+ & 0 & \frac{1}{\sqrt{2}}Pk_- & -\sqrt{\frac{2}{3}}Pk_z & -\frac{1}{\sqrt{3}}Pk_z & \frac{1}{\sqrt{3}}Pk_+  
    \\
    
    0 & T_{\rm{c}} & -\sqrt{\frac{2}{3}}Pk_z & -\frac{1}{\sqrt{2}}Pk_+ & 0 &  -\frac{1}{\sqrt{6}}Pk_- & \frac{1}{\sqrt{3}}Pk_- & \frac{1}{\sqrt{3}}Pk_z     
    \\
    
    \frac{1}{\sqrt{6}}Pk_- & -\sqrt{\frac{2}{3}}Pk_z & T_{\rm{lh}} & -\Omega_2^\dagger & \Omega_1 & 0 & \sqrt{\frac{3}{2}}\Omega_2 & -\sqrt{2}\Omega_3
    \\
    
    0 & -\frac{1}{\sqrt{2}}Pk_- & -\Omega_2 & T_{\rm{hh}} & 0 & \Omega_1 & -\sqrt{2}\Omega_1^\dagger & \frac{1}{\sqrt{2}}\Omega_2 
    \\
    
    \frac{1}{\sqrt{2}}Pk_+ & 0 & \Omega_1^\dagger & 0 & T_{\rm{hh}} & \Omega_2^\dagger & \frac{1}{\sqrt{2}}\Omega_2^\dagger & \sqrt{2}\Omega_1^\dagger  
    \\

    -\sqrt{\frac{2}{3}}Pk_z & -\frac{1}{\sqrt{6}}Pk_+ & 0 & \Omega_1^\dagger & \Omega_2 & T_{\rm{lh}} & \sqrt{2}\Omega_3 & \sqrt{\frac{3}{2}}\Omega_2^\dagger  
    \\

    -\frac{1}{\sqrt{3}}Pk_z & \frac{1}{\sqrt{3}}Pk_+ & \sqrt{\frac{3}{2}}\Omega_2^\dagger & -\sqrt{2}\Omega_1 & \frac{1}{\sqrt{2}}\Omega_2 & \sqrt{2}\Omega_3 & T_{\rm{soff}} & 0  
    \\
    
    \frac{1}{\sqrt{3}}Pk_- & \frac{1}{\sqrt{3}}Pk_z & -\sqrt{2}\Omega_3 & \frac{1}{\sqrt{2}}\Omega_2^\dagger & \sqrt{2}\Omega_1 & \sqrt{\frac{2}{3}}\Omega_2 & 0 & T_{\rm{soff}}  
    \\
    
  \end{pmatrix},
\end{equation} 
where the diagonal terms are
\begin{eqnarray}
T_{\rm{c}}=E_c+\frac{\hbar^2\vec{k}^2}{2m_e}-e\phi(\vec{r}), \\
\label{ZBonsites}
T_{\rm{lh}}=E_h+\Omega_0^{\rm{lh}}-e\phi(\vec{r}),\; \Omega_0^{\rm{lh}}=\frac{\hbar^2}{2m_e}\left[(k_x^2+k_y^2)(\gamma_3-\gamma_1)-k_z^2(2\gamma_3+\gamma_1)\right],  \\
T_{\rm{hh}}=E_h+\Omega_0^{\rm{hh}}-e\phi(\vec{r}),\;\Omega_0^{\rm{hh}}=\frac{\hbar^2}{2m_e}\left[-(k_x^2+k_y^2)(\gamma_3+\gamma_1)+k_z^2(2\gamma_3-\gamma_1)\right],  \\
T_{\rm{soff}}=E_{\rm{soff}}+\Omega_0^{\rm{soff}}-e\phi(\vec{r}),\;\Omega_0^{\rm{soff}}=-\gamma_1\frac{\hbar^2\vec{k}^2}{2m_e},
\end{eqnarray}
and the off-diagonal ones are
\begin{eqnarray}
\Omega_1=-\frac{1}{\sqrt{3}}\frac{\hbar^2}{2m_e}(\gamma_2+2\gamma_3)k_-^2,\\
\Omega_2=-\frac{2}{\sqrt{3}}\frac{\hbar^2}{2m_e}(2\gamma_2+\gamma_3)k_zk_-, \\
\label{ZBcouplings}
\Omega_3=\frac{\hbar^2}{2m_e}\gamma_3 (k_x^2+k_y^2-2k_z^2).
\end{eqnarray}
Here, $E_c$, $E_h=E_c-\Delta_g$ and $E_{\rm{soff}}=E_c-\Delta_g-\Delta_{\rm{soff}}$ are the band edges of the conduction, hole and split-off bands, respectively; $\Delta_g$ and $\Delta_{\rm{soff}}$ are the gaps between the conduction/hole and hole/split-off bands at the $\Gamma$ point; $\phi(\vec{r})$ is the electrostatic potential; $k_\pm\equiv k_x\pm ik_y$ and $\left\lbrace\gamma_i\right\rbrace$ and $P$ are Kane parameters. In this work, we choose the conduction band edge as the reference energy, i.e., we fix $E_c=0$. The Hamiltonian elements whose functional form has been substituted by phenomenological parameters are
\begin{eqnarray}
P\equiv-\frac{i\hbar}{m_e}\left<S\right|p_x\left|X\right>=-\frac{i\hbar}{m_e}\left<S\right|p_y\left|Y\right>=-\frac{i\hbar}{m_e}\left<S\right|p_z\left|Z\right>,  \\
\Delta_{\rm{soff}}\equiv\frac{3\hbar i}{4m_e^2c^2}\left<X\right|\frac{\partial \phi}{\partial x}p_y-\frac{\partial \phi}{\partial y}p_x\left|Y\right>=
\frac{3\hbar i}{4m_e^2c^2}\left<Y\right|\frac{\partial \phi}{\partial y}p_z-\frac{\partial \phi}{\partial z}p_y\left|Z\right>= \frac{3\hbar i}{4m_e^2c^2}\left<Z\right|\frac{\partial \phi}{\partial z}p_x-\frac{\partial \phi}{\partial x}p_z\left|X\right>.
\end{eqnarray}
In order to avoid the spurious solutions coming from the loss of ellipticity of the Schr\"odinger equation \cite{Burt:IOP92, Foreman:PRB97, Veprek:PRB07}, the Kane parameters $\left\{\gamma_i\right\}$ are renormalized from the Luttinger ones $\left\{\gamma_i^{(L)}\right\}$,
\begin{eqnarray}
\left\{ \begin{array}{l}
\gamma_1=\gamma_1^{(L)}-\frac{E_p}{3\Delta_g} \\
\gamma_2=\gamma_2^{(L)}-\frac{E_p}{6\Delta_g} \\
\gamma_3=\gamma_3^{(L)}-\frac{E_p}{6\Delta_g},
\end{array}
   \right.
\end{eqnarray}
where $E_p=\frac{2m_e}{\hbar^2}P^2$ is the Kane energy. The specific values that we have used  for the Kane parameters are extracted from Ref.  \onlinecite{Vurgaftman:JAP01} and shown in Table \ref{Table2}.

\setlength{\tabcolsep}{20pt}
\begin{table}[h]
Table \ref{Table2}: Band and Kane/Luttinger parameters for 8-band Zinc-blende Hamiltonians, extracted from Ref. \onlinecite{Vurgaftman:JAP01}.
\label{Table2}
\begin{tabular}{|c|cccc|}
\hline
 Parameter & InAs & InSb & GaAs & GaSb \\ \hline \hline
$\Delta_g$ (meV) & 417 & 235 & 1519 & 812 \\
$\Delta_{\rm{soff}}$ (meV) & 390 & 810& 341 & 760 \\
$P$ (meV$\cdot$nm) & 919.7 & 940.2 & 1047.5 & 971.3 \\
$\gamma_1^{(L)}$ & 20.0 & 34.8 & 6.98 & 13.4 \\
$\gamma_2^{(L)}$ & 8.5 & 15.5 & 2.06 & 4.7 \\
$\gamma_3^{(L)}$ & 9.2 & 16.5 & 2.93 & 6.0 \\ \hline
\end{tabular}
\end{table}

\subsection{8-band Wurtzite Hamiltonian}

The 8B k$\cdot$p Kane Hamiltonian for Wurtzite crystals has been derived and described in previous works \cite{Faria:PRB16}. The basis for Wurtzite-type semiconductors is given by
\begin{eqnarray}
\left\{ \begin{array}{ll}
\left| C_{\uparrow} \right>=i\left| S_{\uparrow} \right>,  &
\left| C_{\downarrow} \right>=i\left| S_{\downarrow} \right>, \\
\left| LH_{\uparrow} \right>=\frac{1}{\sqrt{2}}\left| (X-iY)_{\uparrow} \right>, &
\left| LH_{\downarrow} \right>=-\frac{1}{\sqrt{2}}\left| (X+iY)_{\downarrow}\right>, \\
\left| HH_{\uparrow} \right>=-\frac{1}{\sqrt{2}}\left| (X+iY)_{\downarrow} \right>, &
\left| HH_{\downarrow} \right>=\frac{1}{\sqrt{2}}\left| (X-iY)_{\uparrow} \right>, \\
\left| SO_{\uparrow} \right>=\left| Z_{\uparrow}\right>, &
\left| SO_{\downarrow} \right>=\left| Z_{\downarrow} \right>,
\end{array}
   \right.
\end{eqnarray}
where the $z$-direction is taken along the $(0001)$ crystallographic orientation.
Its corresponding Hamiltonian in the $\Psi=(\Psi_{\rm{hh},\uparrow},\Psi_{\rm{lh},\uparrow},\Psi_{\rm{soff},\uparrow},\Psi_{\rm{hh},\downarrow},\Psi_{\rm{lh},\downarrow},\Psi_{\rm{soff},\downarrow},\Psi_{\rm{c},\uparrow},\Psi_{\rm{c},\downarrow})$ basis is
\begin{eqnarray}
\label{WHamiltonian}
H_{\rm{kp}}= \left(
  \begin{matrix}
    \Delta_{1}+\Delta_2+\lambda+\theta+E_h-e\phi & -K^{\dagger} & i\left(\alpha_4-\frac{\alpha_1}{\sqrt{2}}\right)k_- -H^{\dagger} & 0
    \\
    
    -K & \Delta_{1}-\Delta_2+\lambda+\theta+E_h-e\phi & -i\left(\alpha_4+\frac{\alpha_1}{\sqrt{2}}\right)k_+ +H & -i\alpha_2k_-
    \\
    
    -i\left(\alpha_4-\frac{\alpha_1}{\sqrt{2}}\right)k_+-H & i\left(\alpha_4+\frac{\alpha_1}{\sqrt{2}}\right)k_- +H^{\dagger} & \lambda+E_h-e\phi & 0
    \\
    
    0 & -\frac{1}{\sqrt{2}}Pk_- & 0 & \Delta_{1}+\Delta_2+\lambda+\theta+E_h -e\phi 
    \\
    
    i\alpha_2k_+ & 0 & \sqrt{2}\Delta_3+i\sqrt{2}\alpha_1k_z & -K^{\dagger}
    \\
    
    0 & \sqrt{2}\Delta_3-i\sqrt{2}\alpha_1k_z & i\alpha_3k_+ & i\left(\alpha_4-\frac{\alpha_1}{\sqrt{2}}\right)k_-+H^{\dagger}    
    \\
    
    -\left(\frac{P_2}{\sqrt{2}}-\frac{P_3}{\sqrt{2}}\right)k_+ +T & \left(\frac{P_2}{\sqrt{2}}+\frac{P_3}{\sqrt{2}}\right)k_- +T^{\dagger} & P_1k_z+U^{\dagger} & 0 
    
    \\
    
    0 & -i\sqrt{2}\Delta_4-\sqrt{2}P_3 k_z & P_4 k_+ & \left(\frac{P_2}{\sqrt{2}}-\frac{P_3}{\sqrt{2}}\right)k_- +T^{\dagger}
    
    \\
  \end{matrix}  \right.
  \nonumber \\
    \nonumber \\
  \left.
  \begin{matrix} 
    ... & -i\alpha_2k_- & 0 & -\left(\frac{P_2}{\sqrt{2}}-\frac{P_3}{\sqrt{2}}\right)k_- +T^{\dagger} & 0
    \\
    
    ... & 0 & \sqrt{2}\Delta_3+i\sqrt{2}\alpha_1k_z & \left(\frac{P_2}{\sqrt{2}}+\frac{P_3}{\sqrt{2}}\right)k_+ +T & i\sqrt{2}\Delta_4-\sqrt{2}P_3 k_z      
    \\
    
    ... & \sqrt{2}\Delta_3-i\sqrt{2}\alpha_1k_z & -i\alpha_3k_- & P_1k_z+U & P_4 k_- 
    \\
    
    ... & -K & -i\left(\alpha_4-\frac{\alpha_1}{\sqrt{2}}\right)k_+ +H & 0 & \left(\frac{P_2}{\sqrt{2}}-\frac{P_3}{\sqrt{2}}\right)k_+ +T 
    \\
    ... & \Delta_{1}-\Delta_2+\lambda+\theta+E_h-e\phi & i\left(\alpha_4+\frac{\alpha_1}{\sqrt{2}}\right)k_- -H^{\dagger}  & i\sqrt{2}\Delta_4-\sqrt{2}P_3 k_z & -\left(\frac{P_2}{\sqrt{2}}+\frac{P_3}{\sqrt{2}}\right)k_- +T^{\dagger} 
    \\
    
    ... & i\left(\alpha_4+\frac{\alpha_1}{\sqrt{2}}\right)k_+ -H & \lambda+E_h-e\phi & -P_4 k_+ & P_1k_z+U      
    \\
    
    ... & -i\sqrt{2}\Delta_4-\sqrt{2}P_3 k_z & -P_4 k_- & E_c+V-e\phi & -i\alpha_5 k_- 
    \\
    
    ... & -\left(\frac{P_2}{\sqrt{2}}+\frac{P_3}{\sqrt{2}}\right)k_+ +T  & P_1k_z+U^{\dagger} & i\alpha_5 k_+ & E_c+V-e\phi

  \end{matrix}
  \right), \ \ \ \ \ \ \ \ \ \ \ \ 
\end{eqnarray}
where
\begin{eqnarray}
\lambda \equiv \gamma_1k_z^2+\gamma_2(k_x^2+k_y^2), \\
\theta \equiv \gamma_3k_z^2+\gamma_4(k_x^2+k_y^2), \\
K=\gamma_5k_+^2, \\
H=\gamma_6k_+k_z, \\
U \equiv i\left[\tilde{P}_1k_z^2+\tilde{P}_2(k_x^2+k_y^2)\right],\\
T\equiv i\tilde{P}_3k_+ k_z, \\
V\equiv e_1k_z^2+e_2(k_x^2+k_y^2).
\end{eqnarray}
$\Delta_i$, $e_i$, $\alpha_i$, $P_i$, $\tilde{P}_i$ and $\gamma_i$ are the band and Kane paremeters. As before, $E_h$ and $E_c$ are the hole and conduction band edges. The explicit form of these parameters is provided in Ref. \cite{Faria:PRB16}. The specific values we use are extracted from Ref. \onlinecite{Faria:PRB16} and shown in Table \ref{Table3}. We choose the conduction band edge as the reference energy, fixing $E_c=0$.

\begin{table}[h!]
Table \ref{Table3}: Band and Kane parameters for the 8-band Wurtzite InAs Hamiltonian, extracted from Ref. \onlinecite{Faria:PRB16}. The parameters $\Delta_g$ and $\Delta_{\rm{soff}}$ are obtained as a combination of $\Delta_{i}$ and $E_c,E_h$, see Ref. \cite{Faria:PRB16}.
\label{Table3}
\begin{tabular}{|ll|ll|ll|}
\hline
\multicolumn{6}{|c|}{InAs} \\ \hline \hline
\multicolumn{2}{|c|}{\begin{tabular}[c]{@{}c@{}}Energy splittings\\ (meV)\end{tabular}} & \multicolumn{2}{|c|}{\begin{tabular}[c]{@{}c@{}}Linear parameters\\ (meV$\cdot$nm)\end{tabular}} & \multicolumn{2}{|c|}{\begin{tabular}[c]{@{}c@{}}Second order parameters\\ (in units of $\hbar^2/2m_e$)\end{tabular}} \\  \hline
$\Delta_1$ & 100.3 & $P_1$ & 838.6 & $\gamma_1$ & 1.5726 \\
$\Delta_2$ & 102.3 & $P_2$ & 689.87 & $\gamma_2$ & -1.6521 \\
$\Delta_3$ & 104.1 & $P_3$ & -6.95 & $\gamma_3$ & -2.6301 \\
$\Delta_4$ & 38.8 & $P_4$ & -21.71 & $\gamma_4$ & 0.5126 \\
$E_c$ & 0 & $\alpha_1$ & -1.89 & $\gamma_5$ & 0.1172 \\
$E_h$ & -664.9 & $\alpha_2$ & -28.92 & $\gamma_6$ & 1.3103 \\
 \multicolumn{2}{|c|}{\hrulefill} & $\alpha_3$ & -51.17  & $\tilde{P}_1$ & -2.3925 \\
 $\Delta_g$ & 467  & $\alpha_4$ & -49.04 & $\tilde{P}_2$ & 2.3155 \\
 $\Delta_{\mathrm{soff}}$ & 325.7 & $\alpha_5 $ & 53.06 & $\tilde{P}_3$ & -1.7231 \\
 &  &  &  & $e_1$ & -3.2005 \\
 &  &  &  & $e_2$ & 0.6363 \\ \hline
\end{tabular}
\end{table}

\section{Derivation of the conduction band approximation}
\label{App:B}

The multiband Hamiltonians presented before correctly describe the band shape of III-V semiconductors around the $\Gamma$-point \cite{Winkler:03, Niquet:PRB06, Kishore:IOP12, Ehrhardt:14, Soluyanov:PRB16, Luo:AIP16, Campos:PRB18}. However, dealing with these Hamiltonians in certain situations is a difficult task due to the large number of bands involved \cite{Ehrhardt:14}.
When one is interested only on the properties of one specific band, the conduction band in this case, it is customary to look for an effective Hamiltonian within that band by integrating out the rest of them. To do this, we start from the $8\times8$ Zinc-blende Hamiltonian of Eq. \eqref{ZBHamiltonian}, which can be written as
\begin{eqnarray}
H_{\rm{kp}}=\begin{pmatrix}
H_{\rm{c}} & H_{\rm{cv}} \\
H_{\rm{cv}}^\dagger & H_{\rm{v}}
\end{pmatrix},
\end{eqnarray}
where $H_{\rm{c}}$ is a $2\times 2$ Hamiltonian corresponding to the (spinful) conduction band, $H_{\rm{v}}$ is a $6\times 6$ Hamiltonian corresponding to the three (spinful) valence bands, and $H_{\rm{cv}}$ is a $2\times 6$ Hamiltonian that represents the coupling between conduction and valence bands. Following a standard folding-down procedure, we can write an effective $2\times 2$ conduction band Hamiltonian as
\begin{eqnarray}
H_{\rm{CB}}=H_{\rm{c}}+\Sigma_{\rm{v}},\;\; \Sigma_{\rm{v}}=H_{\rm{cv}}G_{\rm{v}} H_{\rm{cv}}^\dagger,\;\;G_{\rm{v}}=(E-H_{\rm{v}})^{-1}\;\;\rightarrow H_{\rm{CB}}\Psi_{\rm{c}}=E\Psi_{\rm{c}},
\end{eqnarray}
where $\Sigma_{\rm{v}}$ is the self-energy of the valence bands and $G_{\rm{v}}$ their corresponding Green's function (resolvent).

The valence bands self-energy cannot be found exactly, since it is not possible to invert a matrix that has non-commuting terms such as the position-dependent electrostatic potential, $\phi(\vec{r})$, and the momentum operators $\vec{k}$. 
We can nevertheless expand $G_{\rm{v}}$ in terms of the small parameter $\sim\Omega_i/\Delta_i$, where $\Omega_i=\hbar^2\vec{k}^2\gamma_i/(2 m_e)$ represents the energy associated to the valence band couplings, and $\Delta_i$ are the different band gap energies at the $\Gamma$ point. In particular, $\gamma_i$ represents different combinations of $\gamma_{1,2,3}$, see Eqs. \eqref{ZBonsites}-\eqref{ZBcouplings}. On the other hand, $\Delta_i$ represents either $\Delta_{g}$ or $\Delta_{g}+\Delta_{\rm{soff}}$. The quantitative values of these Kane couplings and band gaps can be found in Table \ref{Table2}. 

Defining
\begin{eqnarray}
H_{\rm{v}}^{(0)}=
  \begin{pmatrix}
    E_h-e\phi(\vec{r}) & 0 & 0 & 0 & 0 & 0
    \\
	0 & E_h-e\phi(\vec{r}) & 0 & 0 & 0 & 0
    \\
    0 & 0 & E_h-e\phi(\vec{r}) & 0 & 0 & 0
    \\
	0 & 0 & 0 & E_h-e\phi(\vec{r}) & 0 & 0
    \\
	0 & 0 & 0 & 0 & E_{\rm{soff}}-e\phi(\vec{r}) & 0
    \\
	0 & 0 & 0 & 0 & 0 & E_{\rm{soff}}-e\phi(\vec{r})
  \end{pmatrix},
\end{eqnarray}
and
\begin{eqnarray}
\label{V}
V=
  \begin{pmatrix}
     \Omega_0^{\mathrm{lh}} & -\Omega_2^\dagger & \Omega_1 & 0 & \sqrt{\frac{3}{2}}\Omega_2 & -\sqrt{2}\Omega_3
    \\    
    -\Omega_2 & \Omega_0^{\mathrm{hh}} & 0 & \Omega_1 & -\sqrt{2}\Omega_1^\dagger & \frac{1}{\sqrt{2}}\Omega_2 
    \\    
    \Omega_1^\dagger & 0 & \Omega_0^{\mathrm{hh}} & \Omega_2^\dagger & \frac{1}{\sqrt{2}}\Omega_2^\dagger & \sqrt{2}\Omega_1^\dagger  
    \\
    0 & \Omega_1^\dagger & \Omega_2 & \Omega_0^{\mathrm{lh}} & \sqrt{2}\Omega_3 & \sqrt{\frac{3}{2}}\Omega_2^\dagger  
    \\
    \sqrt{\frac{3}{2}}\Omega_2^\dagger & -\sqrt{2}\Omega_1 & \frac{1}{\sqrt{2}}\Omega_2 & \sqrt{2}\Omega_3 & \Omega_0^{\mathrm{soff}} & 0  
    \\    
    -\sqrt{2}\Omega_3 & \frac{1}{\sqrt{2}}\Omega_2^\dagger & \sqrt{2}\Omega_1 & \sqrt{\frac{2}{3}}\Omega_2 & 0 & \Omega_0^{\mathrm{soff}}  
    \\    
  \end{pmatrix},
\end{eqnarray}
we can write the zeroth-order valence bands Green's function as $G_{\rm{v}}^{(0)}=(E-H_{\rm{v}}^{(0)})^{-1}$, and the Dyson expansion of the full Green's function as
\begin{equation}
\label{Dyson}
G_{\rm{v}}=G_{\rm{v}}^{(0)}+G_{\rm{v}}^{(0)}VG_{\rm{v}}^{(0)}+G_{\rm{v}}^{(0)}VG_{\rm{v}}^{(0)}VG_{\rm{v}}^{(0)}+...
\end{equation}

Ignoring the couplings between the valence bands (while still retaining the couplings between the conduction and valence bands), i.e., to zeroth-order in $\Omega_i/\Delta_i$, the zeroth-order conduction band Hamiltonian 
\begin{eqnarray}
H_{\rm{CB}}^{(0)}=H_{\rm{c}}+H_{\rm{cv}}G_{\rm{v}}^{(0)}H_{\rm{cv}}^\dagger\rightarrow H_{\rm{CB}}^{(0)}\Psi_{\rm{c}}^{(0)}=E\Psi_{\rm{c}}^{(0)},
\end{eqnarray}
has been found \cite{Darnhofer:PRB93, Winkler:03, Wojcik:PRB18} to be
\begin{eqnarray}
\label{HLambdas}
H_{\rm{CB}}^{(0)}=\left[\frac{\hbar^2\vec{k}^2}{2m_e}+E_c-e\phi(\vec{r})\right]\sigma_0+(\Lambda_0\sigma_0+\Lambda_x\sigma_x+\Lambda_y\sigma_y+ \Lambda_z\sigma_z),
\end{eqnarray}
where $\sigma_i$ are the spin Pauli matrices (and $\sigma_0$ the identity), and
\begin{eqnarray}
\label{Lambdas}
\Lambda_0=-\sum_{j=\left\{x,y,z\right\}}\frac{P^2}{3}k_j\left[\frac{2}{E_h-e\phi(\vec{r})-E}+\frac{1}{E_{\rm{soff}}-e\phi(\vec{r})-E}\right]k_j, \ \ \ \\
\Lambda_x=i\frac{P^2}{3}\left\{k_z\left[\frac{1}{E_h-e\phi(\vec{r})-E}-\frac{1}{E_{\rm{soff}}-e\phi(\vec{r})-E}\right]k_y-k_y\left[\frac{1}{E_h-e\phi(\vec{r})-E}-\frac{1}{E_{\rm{soff}}-e\phi(\vec{r})-E}\right]k_z\right\}, \ \ \ \\
\Lambda_y=i\frac{P^2}{3}\left\{k_x\left[\frac{1}{E_h-e\phi(\vec{r})-E}-\frac{1}{E_{\rm{soff}}-e\phi(\vec{r})-E}\right]k_z-k_z\left[\frac{1}{E_h-e\phi(\vec{r})-E}-\frac{1}{E_{\rm{soff}}-e\phi(\vec{r})-E}\right]k_x\right\}, \ \ \ \\
\Lambda_z=i\frac{P^2}{3}\left\{k_y\left[\frac{1}{E_h-e\phi(\vec{r})-E}-\frac{1}{E_{\rm{soff}}-e\phi(\vec{r})-E}\right]k_x-k_x\left[\frac{1}{E_h-e\phi(\vec{r})-E}-\frac{1}{E_{\rm{soff}}-e\phi(\vec{r})-E}\right]k_y\right\}. \ \ \
\end{eqnarray}
This Hamiltonian can be recasted in the form
\begin{equation}
H_{\rm{CB}}^{(0)}=\left[\vec{k}\frac{\hbar^2}{2m^{(0)}(\vec{r})}\vec{k}+E_c-e\phi(\vec{r})\right]\sigma_0+\frac{1}{2}\left[\vec{\alpha}_R^{(0)}(\vec{r})\cdot \left(\vec{\sigma}\times\vec{k}\right)+\left(\vec{\sigma}\times\vec{k}\right)\cdot\vec{\alpha}^{(0)}_R(\vec{r})\right].
\label{Hrecasted}
\end{equation}
The first term corresponds to the kinetic energy, but the electron has now an effective mass given by
\begin{equation}
\frac{1}{m^{(0)}(\vec{r})}=\frac{1}{m_e}-\frac{2P^2}{3\hbar^2}\left[\frac{2}{E_h-e\phi(\vec{r})-E}+\frac{1}{E_{\rm{soff}}-e\phi(\vec{r})-E}\right].
\label{Eq:mapendix}
\end{equation}
This term has to be written between the two momentum operators $\vec{k}$ in the kinetic energy part of \eqref{Hrecasted} because the effective mass depends in general on position through $\phi(\vec{r})$.
The other term corresponds to the Rashba SO interaction, whose Rashba coefficients $\vec{\alpha}_R^{(0)}(\vec{r})$ are
\begin{eqnarray}
\vec{\alpha}^{(0)}_{R}(\vec{r})=\frac{P^2}{3}\vec{\nabla}\left[\frac{1}{E_h-e\phi(\vec{r})-E}-\frac{1}{E_{\rm{soff}}-e\phi(\vec{r})-E}\right].
\label{Eq:Rashba}
\end{eqnarray}
Note that, in this notation, the nabla operator inside $\vec{\alpha}_R^{(0)}(\vec{r})$ only acts on the expression inside the square brackets, i.e., it is not to be applied on the electron wave function.


Equation \eqref{Hrecasted} is equivalent to Eq. \eqref{Eq:H_cb} of the main text, but taking the conduction band edge $E_c$ as the reference energy, i.e., $E_c=0$. It describes the Hamiltonian of an electron quasiparticle in the conduction band, and therefore, it is only valid as far as conduction and valence bands do not overlap in energy, what means that its applicability range is for $-E_h=\Delta_g-E_c>-e\phi(r)-E$. We have thus reached a Hamiltonian that involves a smaller number of bands (i.e., only one spinful conduction band), and that allows us to directly introduce additional terms whose functional form we know, as for instance a Zeeman field or a superconducting pairing term. We note that this simplified Hamiltonian is equivalent to that of the bare electron \footnote{Using the properties of the triple product, we can equate $(\vec{\nabla}\phi\times\vec{k})\cdot\vec{\sigma}=-\vec{\nabla}\phi\cdot(\vec\sigma\times\vec{k})$.}, Eq. \eqref{Eq:Hamiltonian_FreeElectron}, but with a spatial-dependent effective mass $m^{(0)}(\vec{r})$ and an effective Rashba coupling $\vec{\alpha}_R^{(0)}(\vec{r})$ that depends not only on $\vec{\nabla}\phi(\vec{r})$, but whose prefactor, like that of the effective mass, depends on $E+\phi(\vec{r})$, where $E$ is the quasiparticle's energy. Due to the energy dependence of these effective quantities, the conduction band Hamiltonian has to be solved self-consistently.

To avoid this complication, a further simplification is usually performed in the literature \cite{Darnhofer:PRB93, Winkler:03, Wojcik:PRB18}. If $\Delta_g$ and $\Delta_{\rm{soff}}$ are the largest energies in the conduction band approximation, then it is possible to expand in Taylor series assuming $ \left|E_h\right| \gg \left|e\phi(\vec{r})+E\right|$,
\begin{eqnarray}
\frac{1}{E_h-e\phi(\vec{r})-E}=\frac{(-E_h)}{1-\frac{e\phi(\vec{r})+E}{E_h}}=\sum_{n=0}^{\infty}(-1)^{(n+1)}\frac{(e\phi(\vec{r})+E)^n}{(E_h)^{n+1}}.
\end{eqnarray}
Truncating up to $n=0$ in the above equation, it is possible to obtain an energy and position-independent effective mass
\begin{equation}
\frac{1}{m^{(0)}}\simeq\frac{1}{m_e}-\frac{2P^2}{3\hbar^2}\left(\frac{2}{E_h}+\frac{1}{E_{\rm{soff}}}\right),
\label{Eq:Effective_mass}
\end{equation}
which is equivalent to Eq. \eqref{Eq:m_eff-oversimplified} of the main text. Similarly, truncating up to $n=1$, we get the Rashba coupling
\begin{eqnarray}
\vec{\alpha}^{(0)}_{R}(\vec{r})\simeq -e\frac{P^2}{3}\left(\frac{1}{E_h^2}-\frac{1}{E_{\rm{soff}}^2}\right)\vec{\nabla}\phi(\vec{r}),
\label{Eq:Effective_Rashba}
\end{eqnarray}
which we have called \emph{simplified} in the main text, see Eq. \eqref{Eq:alpha_R-oversimplified}.  Equation \eqref{Eq:Effective_mass} is commonly used to estimate the effective masses of typical semiconductors, leading to the widely used values \cite{Vurgaftman:JAP01} of $m\simeq0.023m_e$ for InAs and $m\simeq0.014m_e$ for InSb semiconductors, see App. \ref{App:G} for a more profound discussion. 

If we apply the previous expressions to a semiconductor with the form of a nanowire that is translational invariant along the $z$-direction, as in the main text, we can take $k_z$ as a good quantum number and thus $\phi(\vec{r})$ depends only on $x,y$. In this particular case $\alpha_{R,z}$=0 (to all orders), and we can write the Rashba term in the conduction band Hamiltonian $H_{\rm{CB}}^{(0)}$ as
\begin{eqnarray}
H_R^{(0)}=\frac{1}{2}\left[\alpha_{R,x}^{(0)}(x,y)(\sigma_yk_z-\sigma_zk_y)+\alpha_{R,y}^{(0)}(x,y)(\sigma_zk_x-\sigma_xk_z) \right. \ \ \ \ \ \nonumber \\ 
\left. +(\sigma_yk_z-\sigma_zk_y)\alpha_{R,x}^{(0)}(x,y)+(\sigma_zk_x-\sigma_xk_z)\alpha_{R,y}^{(0)}(x,y)\right].
\end{eqnarray}
The SO coupling modulus thus involves only the $x$ and $y$ components: $|\alpha_R|=\sqrt{\alpha_{R,x}^2+\alpha_{R,y}^2}$.

All the previous analysis of the conduction band approximation has been carried out starting from the 8B Zinc-blende Hamiltonian of Eq.  \eqref{ZBHamiltonian}. A similar derivation could be performed starting from the Wurtzite Hamiltonian of Eq. \ref{WHamiltonian} for the particular case of InAs semiconductors, but this derivation would certainly be more involved due to the presence of more coupling parameters. In practice, the Zinc-blende zeroth-order results for the effective mass and SO coupling derived above are also used for Wurtzite InAs, but taking the values of $\Delta_g$ and $\Delta_{\rm{soff}}$ of Table \ref{Table3} instead of those of Table \ref{Table2}, and taking the Zinc-blende $P$ parameter. In any case, as we discuss in the main text, the zeroth-order SO coupling cannot account for the specific crystal structure, be it Zinc-blende or Wurtzite, when confinement effects are important, since the neglected terms proportional to $\gamma_i$ are essential.

We could go to higher orders in the Dyson expansion of the valence bands Green's function of Eq. \eqref{Dyson}. To find analytical manageable expressions is nevertheless cumbersome, especially as the series order increases. For example, for a Zinc-blende nanowire along the $z$-direction we have found an approximation for the SO coupling to first order given by
\begin{eqnarray}
\vec{\alpha}^{(1)}_{R}(\vec{r})\approx e\frac{2P^2}{3}\left[\frac{\gamma_1-\gamma_3}{(\Delta_g+e\phi(\vec{r})+E)^3}-\frac{\gamma_1}{(\Delta_g+\Delta_{\rm{soff}}+e\phi(\vec{r})+E)^3}\right]\vec{\nabla}\phi(\vec{r})\frac{\hbar^2}{2m_e}(\partial_x^2+\partial_y^2),
\label{Eq:Rashbaorder1}
\end{eqnarray}
where $E_c$ is taken zero like in the main text. This coupling gives a Rashba correction $H_R^{(1)}= [\vec{\alpha}_R^{(1)}(\vec{r})\cdot (\vec{\sigma}\times\vec{k})+(\vec{\sigma}\times\vec{k})\cdot\vec{\alpha}^{(1)}_R(\vec{r})]/2$ to the zeroth-order reduced Hamiltonian of Eq. \eqref{Hrecasted}.  To get to this analytical expression we've performed a couple of approximations. On the one hand, we have realized that in the Hamiltonian of $V$, Eq. \eqref{V}, the relevant terms are the ones in the diagonal, $\Omega_0^{i}$, while we can neglect the off-diagonal couplings $\Omega_{i=\{1,2,3\}}$. We have checked this numerically (not shown). The reason is that when $\Omega_0^{i}$ are different from zero at the $\Gamma$ point, i.e., for transverse subbands with $k_{x}\neq 0$ and/or $k_{y}\neq 0$, they modify the positions of the different valence subbands at $\Gamma$ and thus the value of the effective gaps between them. This in turn contributes significantly to the SO coupling. On the other hand, in Eq. \eqref{Eq:Rashbaorder1} we have ignored terms proportional to $k_z^2$ and higher orders in $\vec{\nabla}\phi(\vec{r})$. Note that to such order, $\vec{\alpha}^{(1)}_{R}$ is Hermitian.

The SO coupling to first order in Eq. \eqref{Eq:Rashbaorder1} is then  proportional to the transverse energy $\hbar^2(k_x^2+k_y^2)/(2m_e)$. This term is zero for an energy subband with zero transverse momentum. However, in a finite-width nanowire with many finite $k_{x,y}\sim 1/W_{\rm{wire}}$ momenta, the contribution given by Eq. \eqref{Eq:Rashbaorder1} or higher order terms can be non-negligible, as we show now.

\begin{figure}
\includegraphics[scale=0.29]{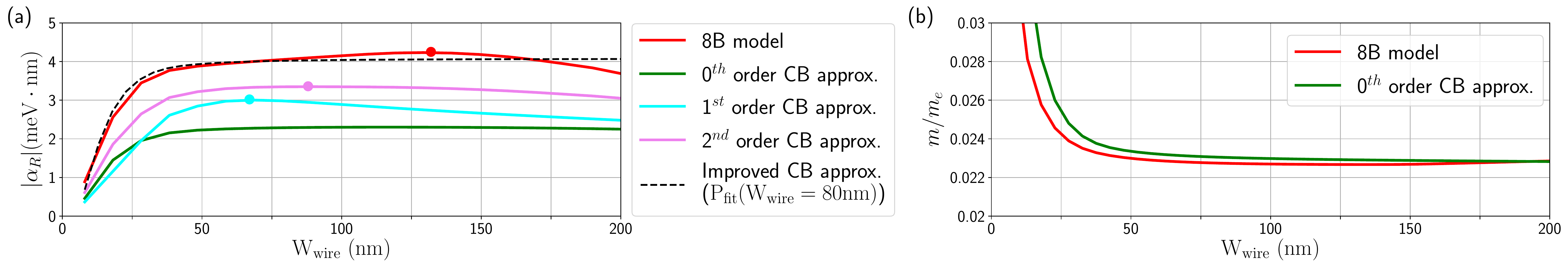}
\caption{(a) Modulus of the SO coupling as a function of the wire's width $W_{\rm{wire}}$ for a Zinc-blende InAs nanowire in a homogeneous electric field $\vec{E}=(0,2,0)$ meV$\cdot$nm. Different degrees of approximation are compared in this figure. In red we show the SO coupling found with the 8B model Hamiltonian of Eq. \eqref{ZBHamiltonian}, considered as the exact result in this work. The result to zeroth-order in the conduction band approximation is shown in green, to first order in cyan and to second order in pink. Dots mark the maximum of each curve. The dashed black curve corresponds to the improved SO Eq. \eqref{Eq:alpha_R_improved}, which considers a constant $P_{\rm{fit}}$ parameter fixed to $W_{\rm{wire}}=80$ nm. (b) Effective electron mass $m$ over bare electron mass $m_e$ versus $W_{\rm{wire}}$. In red the 8B model result and in green the value of the zeroth-order conduction band approximation.}
\label{FigB}
\end{figure}

Let us consider, for instance, a Zinc-blende InAs nanowire subject, for simplicity, to a position-independent electric field in the $y$-direction, particularly $\vec{E}=(0,2,0)$ meV$\cdot$nm. We want to compare the SO coupling calculated to different orders in $\Omega_i/\Delta_i$. In Fig. \ref{FigB}(a) we plot the modulus of the SO coupling as a function of the nanowire's width $W_{\rm{wire}}$. In red we show the result found with the 8B Hamiltonian of Eq. \eqref{ZBHamiltonian}, which is a computationally demanding calculation, especially as $W_{\rm{wire}}$ increases. We consider the red curve --the one provided by the 8B model-- as the exact or correct result in our work. One could get more accurate results by going to higher orders in the L\"owdin perturbation theory explained in App. \ref{App:A} or by going away from the k$\cdot$p theory, but that is beyond the scope of this paper. We see that the SO coupling rapidly increases for small wire widths and then approximately plateaus from $W_{\rm{wire}}\approx 25$ nm to $W_{\rm{wire}}\approx 175$ nm. From that moment on it decreases. In green we have the zeroth-order conduction band approximation calculated from Eq. \eqref{Eq:Rashba} or, equivalently, from Eq. \eqref{Eq:alpha_R} of the main text. It also increases for small widths, for approximately the same $W_{\rm{wire}}$-range as the 8B model result, but then it remains constant and about half the value of the 8B model. In particular, for $W_{\rm{wire}}=80$ nm, it is a $54\%$ of the exact value. We have also worked out numerically the SO coupling to first order, cyan curve, and to second order, pink curve. At $W_{\rm{wire}}=80$ nm again, the cyan curve provides the $75\%$ of the 8B model result, and the pink one the $85\%$. We observe that, in order to approach the red value, we need to sum many orders (if not all) in the Dyson expansion for the realistic wire widths presented in this figure, something that is neither easy nor practical.

In Fig. \ref{FigB} we only get to wire widths of 200 nm, but as $W_{\rm{wire}}\rightarrow\infty$, the red curve should approach the green one, the zeroth-order. Note that the first and second orders also decrease with width. This can be understood because, as the width increases, the transverse momenta $k_{x,y}\sim1/W_{\rm{wire}}$ go to zero and so does $\Omega_i=\hbar^2\vec{k}^2\gamma_i/(2 m_e)$. This means that for bulk semiconductors, the zeroth-order SO coupling commonly found in the literature is a very good approximation. Moreover, for $W_{\rm{wire}}\rightarrow\infty$, the SO coupling of Zinc-blende and Wurtzite InAs are very similar. However, when confinement effects are important, $\alpha_R^{(0)}$ cannot provide the quantitative and the qualitative behavior of the SO coupling, nor distinguish between different crystal structures.

Notice that, although the 8B model SO coupling has a pretty constant value from $W_{\rm{wire}}\approx 25$ nm to $175$ nm, as mentioned above, it has a maximum at $W_{\rm{wire}}=130$ nm. This maximum value depends on the band and Kane parameters in an intricate way. The cyan and pink curves have also a maximum, but for smaller $W_{\rm{wire}}$. Knowing the wire's width for which the SO coupling is maximum can help design semiconductor nanowires with stronger SO effects.

In Fig. \ref{FigB}(b) we study the conduction band effective mass as a function of $W_{\rm{wire}}$. As before, the red curve corresponds to the Zinc-blende InAs 8B model result, whereas the green one to the zeroth-order approximation of Eq. \eqref{Eq:mapendix} or, equivalently, Eq. \eqref{Eq:m_eff} of the main text. Strikingly, we observe that $m^{(0)}$ provides an excellent approximation to the exact red result for basically all wire widths, except for the smallest ones, where, in any case, the k$\cdot$p model is not valid. This means that, while higher corrections in the conduction band approximation are necessary to correctly account for the SO coupling of semiconducting nanowires, the effective mass can be accurately captured with the common zeroth-order limit. The ultimate reason comes from the relative signs between the $1/\Delta_i$ terms in Eqs. \eqref{Eq:mapendix}, \eqref{Eq:Rashba}. For the effective mass these terms are added, while for the SO coupling they are subtracted.

\section{Numerical methods}
\label{App:C}
In order to solve numerically the eigenvalue problem with the previous Hamiltonians, we use the finite difference method (FDM). For a nanowire that is infinite in the $z$-direction (the growth direction) but finite in the $x-y$ plane, this method consists in discretizing the transverse coordinates with a finite mesh of points $(\left\{x_i\right\},\left\{y_j\right\})$. The momentum operator is substituted by its corresponding derivative, which is then discretized using the central difference method
\begin{eqnarray}
k_x\rightarrow -i\frac{\partial}{\partial x}\rightarrow -i\frac{c_{i+1}^{\dagger}c_i-c_{i-1}^{\dagger}c_i}{2(x_{i+1}-x_{i-1})}, \\
k_x^2\rightarrow -\frac{\partial^2}{\partial x^2}\rightarrow -\frac{c_{i+1}^{\dagger}c_i-2c_{i}^{\dagger}c_i+c_{i-1}^{\dagger}c_i}{(x_{i+1}-x_{i})(x_{i}-x_{i-1})}.
\end{eqnarray}
In the $z$-direction no discretization is needed because $k_z$ is a good quantum number due to translational symmetry.

Notice that the FDM method has been shown \cite{Cartoixa:JAP03, Yang:PRB05} to introduce spurious solutions into the energy spectrum of multiband Hamiltonians. In order to alleviate the impact of these spurious states, some mechanisms have been proposed. We follow here Refs. \onlinecite{Wu:PRB06} and \onlinecite{Tan:JOP90}, where we use a staggered grid with a finer discretization close to the wire boundaries to suppress the spurious states. However, if the discretization is not homogeneous, the FDM generates non-Hermitian matrices. For instance, for two consecutive points $x_i$ and $x_{i+1}$,
\begin{equation}
\left. \begin{array}{l}
     \frac{\partial f(x_i)}{\partial x}\rightarrow \frac{f(x_{i+1})-f(x_{i-1})}{x_{i+1}-x_{i-1}} \\
	\frac{\partial f(x_{i+1})}{\partial x}\rightarrow \frac{f(x_{i+2})-f(x_{i})}{x_{i+2}-x_{i}}
        \end{array}
   \right\} \rightarrow
    \frac{f(x_{i+1})}{x_{i+1}-x_{i-1}}\neq \frac{f(x_{i})}{x_{i+2}-x_{i}},
\end{equation}
since, in general, for an arbitrary mesh $(x_{i+1}-x_{i-1})\neq (x_{i+2}-x_{i})$. To correct this problem, we follow the ideas of Refs. \onlinecite{Tan:JOP90} and \onlinecite{Pryor:JAP15} and we symmetrize the discretization operator, defined here as $\xi_{ij}\equiv h/|x_i-x_j|$ where $h\equiv\left<x_{i+1}-x_i\right>$. This is needed because the mesh spacing $|x_{i+1}-x_i|$ is not just a number but a position-dependent operator. With this symmetrization, the derivatives can be written as
\begin{eqnarray}
k_x\rightarrow -i\frac{\partial}{\partial x}\rightarrow -\frac{i}{4h}\left\{(\xi_{i+1,i}+\xi_{i-1,i})c_{i+1}^{\dagger}c_i-(\xi_{i+1,i}+\xi_{i-1,i})c_{i-1}^{\dagger}c_i \right\},\\
k_x^2\rightarrow -\frac{\partial^2}{\partial x^2}\rightarrow -\frac{1}{2h^2}\left\{(\xi_{i+1,i}+\xi_{i,i})c_{i+1}^{\dagger}c_i-(2\xi_{i,i}+\xi_{i+1,i}+\xi_{i-1,i})c_{i}^{\dagger}c_i +(\xi_{i,i}+\xi_{i-1,i})c_{i-1}^{\dagger}c_i \right\},
\end{eqnarray}
which are symmetric operators now.

Once the Hamiltonian is discretized in the transverse cross-section (taking $h=1$ nm in this work), it is diagonalized for each $k_z$ value in order to obtain the band structure $E^{(j)}_{\pm}(k_z)$, with the corresponding eigenfunction spinors denoted by $\Psi^{(j)}_{\pm}(x,y,k_z)$. Here, $j$ denotes a transverse subband index and $\pm$ the two associated spin textures related by time reversal symmetry at $k_z=0$. To obtain these we use the standard ARPACK tools provided by the Python package \textit{Scipy}. 

For the conduction band model, we can split the Hamiltonian of Eq. \eqref{Hrecasted} into transverse and longitudinal parts, 
\begin{eqnarray}
H_{\rm{CB}}^{(0)}=H_T(x,y)+H_L(k_z), \\
H_T(x,y)=\left[ -\partial_x\frac{\hbar^2}{2m^{(0)}(x,y)}\partial_x-\partial_y\frac{\hbar^2}{2m^{(0)}(x,y)}\partial_y+E_c-e\phi(x,y)\right]\sigma_0 \nonumber \\
+\frac{1}{2}\left[\alpha_{R,y}^{(0)}(x,y) k_x- \alpha_{R,x}^{(0)}(x,y) k_y+k_x \alpha_{R,y}^{(0)}(x,y)- k_y \alpha_{R,x}^{(0)}(x,y) \right]\sigma_z,\\
H_L(x,y,k_z)=\frac{\hbar^2k_z^2}{2m^{(0)}(x,y)}+\left[{\alpha}_{R,x}^{(0)}(x,y)\sigma_y-{\alpha}_{R,y}^{(0)}(x,y)\sigma_x\right]k_z,
\end{eqnarray}
where $\alpha_{R,z}^{(0)}=0$ because the nanowire is translationally invariant along the $z$-direction. Now, we assume that the SO length is larger or comparable to the wire's diameter, i.e., $l_{\rm{SO}}\gtrsim W_{\rm{wire}}$, as is the case of every experiment analyzed in this work. This assumption allows us to write the total wave function as $\Psi^{(j)}_{\pm}(x,y,k_z)\approx \Psi^{(j)}_{\pm,T}(x,y)e^{ik_z z}$, which amounts to neglecting the inter-subband SO coupling \cite{Scheid:IOP09, Diez:PRB12, Vuik:IOP16}. 
Hence, we can write the total energy for each transverse subband $j$ as a function of $k_z$ as
\begin{equation}
E^{(j)}_\pm(k_z)=E_T^{(j)}+\frac{\hbar^2k_z^2}{2m_{\rm{eff}}^{(j)}}\pm\left|\alpha_{\rm{eff}}^{(j)}k_z\right|,
\end{equation}
where $E_T^{(j)}=\langle\Psi_{\pm,T}^{(j)}|H_T(x,y)|\Psi_{\pm,T}^{(j)}\rangle$ are the Kramers-degenerate eigenvalues of the transverse subband Hamiltonian (which are found numerically by diagonalizing it), and 
\begin{eqnarray}
\pm\alpha_{\rm{eff}}^{(j)}=\left<\Psi_{\pm,T}^{(j)}\right|({\alpha}_{R,x}^{(0)}(x,y)\sigma_y-{\alpha}_{R,y}^{(0)}(x,y)\sigma_x)\left|\Psi_{\pm,T}^{(j)}\right>
\end{eqnarray}
is the projection of the Rashba coupling onto the transverse basis. Note that in the main text we omit the $\pm$ spin quantum number in $\left|\Psi_{\pm,T}^{(j)}\right>$ for simplicity, as there we are only interested in the magnitude of $\alpha_{\rm{eff}}^{(j)}$ and not its sign.

The charge density for the conduction band approximation is simply given by
\begin{equation}
\rho(x,y)= e\sum_{j,\pm} \int\frac{dk_z}{2\pi} |\Psi_{\pm}^{(j)}(x,y, k_z)|^2 f[E^{(j)}_{\pm}(k_z)]=e\sum_{j,\pm} |\Psi_{\pm,T}^{(j)}(x,y)|^2  \int_{-\infty}^{\infty} dE f(E)D_\pm^{(j)}(E) = e\sum_{j,\pm} |\Psi_{\pm,T}^{(j)}(x,y)|^2 n^{(j)},
\end{equation}
where $f(E)$ is the Fermi-Dirac distribution, $n^{(j)}$ is the occupation of subbands $\Psi_{\pm}^{(j)}$ and $D_\pm^{(j)}$ is the corresponding 1D density of states,
\begin{equation}
D_\pm^{(j)}(E)=\frac{1}{\pi}\left(\frac{dk^{(j)}_{z,\pm}(E)}{dE}\right),
\end{equation}
expressed in terms of the fixed-energy $k^{(j)}_{z,\pm}(E)$ of each mode that satisfies $E^{(j)}_\pm(k^{(j)}_{z,\pm}) = E$.

\section{Electrostatic potential and its numerical solution}
\label{App:D}
The electrostatic potential is given by the solution of the Poisson equation
\begin{equation}
\vec{\nabla}\cdot\left(\epsilon(\vec{r})\vec{\nabla}\phi(\vec{r})\right)=\rho(\vec{r}),
\end{equation}
where $\epsilon(\vec{r})$ is the dielectric permittivity, $\phi(\vec{r})$ is the electrostatic potential and $\rho(\vec{r})$ is the charge density inside the wire. The precise experimental setups considered in this work are taken into account through the inhomogeneous dielectric permittivity $\epsilon(\vec{r})$, which we model as a piecewise function that is constant inside each material and has abrupt changes at the interfaces. To find the potential $\phi(\vec{r})$ throughout all space, we fix as boundary conditions the potentials created by the surrounding gates at the gate boundaries, and we use periodic boundary conditions along the $z$-axis. Due to the surface chemistry, there is typically an electron accumulation layer close to the surface of the wire. We model it by introducing an additional source term $\rho_{\rm{surf}}$ to the Poisson equation, consisting in a 1 nm thick positive charge density layer. We solve the Poisson equation with all these ingredients using the Pyhton package \textit{Fenics} \cite{Logg:10, Logg:12} with a mesh discretization of 1 nm.

We note that, since the charge density of the wire $\rho(\vec{r})$ depends on the wavefunction of the Hamiltonian, which in turn depends on the electrostatic potential, the Poisson and Schr\"odinger equations must be solved self-consistently. For this purpose, to obtain the charge density we employ an iterative method that makes use of the so-called Anderson mixing
\begin{equation}
\rho^{(n)}=\beta\tilde{\rho}^{(n)}+(1-\beta)\rho^{(n-1)},
\label{Anderson_mixing}
\end{equation}
where $n$ is a certain iteration step and $\beta$ is the Anderson coefficient. In the first step of this procedure (i.e., $n=0$) we take $\rho^{(0)}=0$ and compute the electrostatic potential of the system. At any other arbitrary step $n$, we compute the charge density $\tilde{\rho}^{(n)}$ using the electrostatic potential found in the previous iteration $n-1$. Then, we compute the electrostatic potential at the $n$-th step using $\rho^{(n)}$, given by the Anderson mixing of Eq. \eqref{Anderson_mixing}. This charge density mixing between the step $n$ and $n-1$ ensures the convergence to the solution. We keep this iterative procedure until the cumulative error is bellow the 1\%. As Anderson coefficient $\beta$, we take a self-adaptive one,
\begin{equation}
\beta=\beta^{(max)}\exp{\left( \frac{\max{(\left||\rho^{(n)}|-|\rho^{(n-1)}|\right|)}}{\max(|\rho^{(n)}|,|\rho^{(n-1)}|)} \right)},
\end{equation}
where $\beta^{(max)}$ is the maximum value the Anderson coefficient can take while ensuring converge. This value depends on the particular system considered.

\section{Improved $P_{\rm{fit}}$ parameter for Zinc-blende InAs, InSb, GaAs and GaSb nanowires}
\label{App:E}
The conduction band approximation and the improved Eq. \eqref{Eq:alpha_R_improved}, whose validity has been proved for InAs in the main text, works also for any other III-V binary compound semiconductor. Among these compounds, InAs, InSb, GaAs and GaSb have the largest Rashba SO coupling because they present the smallest band gap energies.  Therefore, we calculate the improved $P_{\rm{fit}}$ parameter for these materials. In particular, we focus on  Zinc-blende structures since they have larger SO couplings than the Wurtzite counterparts.

\begin{figure}
\includegraphics[scale=0.27]{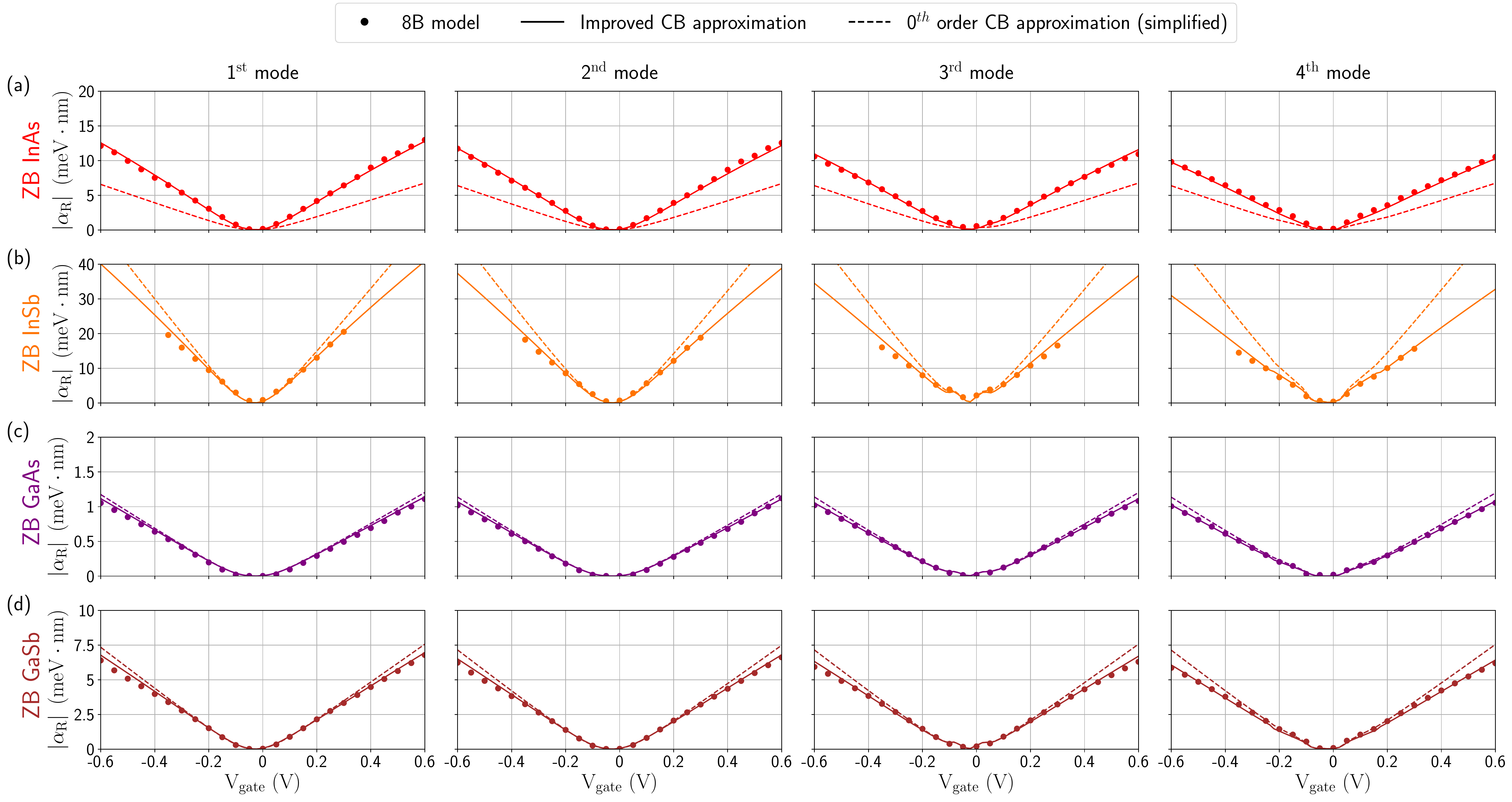}
\caption{Rashba SO coupling modulus versus gate voltage for different Zinc-blende (ZB) (111) nanowires (rows) and different transverse modes (columns). Dots correspond to $\alpha_R$ obtained from the 8B k$\cdot$p Kane model of App. \ref{App:A} using Eq. \eqref{E_fit_kp}. Solid lines correspond to $\alpha_R$ obtained from the improved Eq. \eqref{Eq:alpha_R_improved} within the conduction band approximation discussed in Sec. \ref{Sec-CB_approx}. The corresponding $P_{\rm{fit}}$ parameters are given in Table \ref{Table1}. Dashed lines correspond to $\alpha_R$ obtained in the approximation $\Delta_g\gg \left|e\phi(\vec{r})+E\right|$, i.e., using the simplified Eq. \eqref{Eq:alpha_R-oversimplified}. The simulations have been performed using the same electrostatic environment and parameters as in Fig. \ref{Fig2} of the main text, but changing the dielectric permittivity of the wires to their corresponding ones, $\epsilon_{\rm{InAs}}=15.15$, $\epsilon_{\rm{InSb}}=16.8$, $\epsilon_{\rm{GaAs}}=12.9$ or $\epsilon_{\rm{GaSb}}=15.7$. These values are extracted from Ref. \onlinecite{Levinshtein:00}.}
\label{FigE}
\end{figure}

In Fig. \ref{FigE} we show the magnitude of the SO Rashba coefficients as a function of the back gate potential. Dots are used for the 8B k$\cdot$p simulations. Solid lines correspond to $\alpha_R$ obtained from the improved Eq. \eqref{Eq:alpha_R_improved} discussed in Sec. \ref{Sec-CB_approx}. The corresponding $P_{\rm{fit}}$ parameters are presented in Table \ref{Table1}. These values are extracted by fitting Eq. \eqref{Eq:alpha_R_improved} to the lowest-energy mode of the 8B model simulations, as we do in the main text for InAs. With dashed lines we also plot the result of the simplified  Eq. \eqref{Eq:alpha_R-oversimplified}, for comparison purposes. As can be seen, the improved equation works equally well for every type of semiconductor. However, the simplified equation is a good approximation only for GaAs and GaSb, since these compounds present a large semiconducting gap $\Delta_g$, and therefore the condition $\Delta_g \gg \left|e\phi(\vec{r})+E\right|$ is fulfilled for a wide range of gate potentials.

We note that the material with the largest SO coupling is InSb, which is roughly twice that of InAs, as it was pointed out in previous works \cite{Campos:PRB18, Takase:IOP19}. This is due to the small band gap and large split-off gap that characterizes InSb. In contrast, GaAs exhibits the smallest SO coupling of these materials. This is due to its large band gap and small split-off gap.

\section{Dependence of the improved $P_{\rm{fit}}$ parameter on the electrostatic environment}
\label{App:F}
\begin{figure}[h]
\includegraphics[scale=0.27]{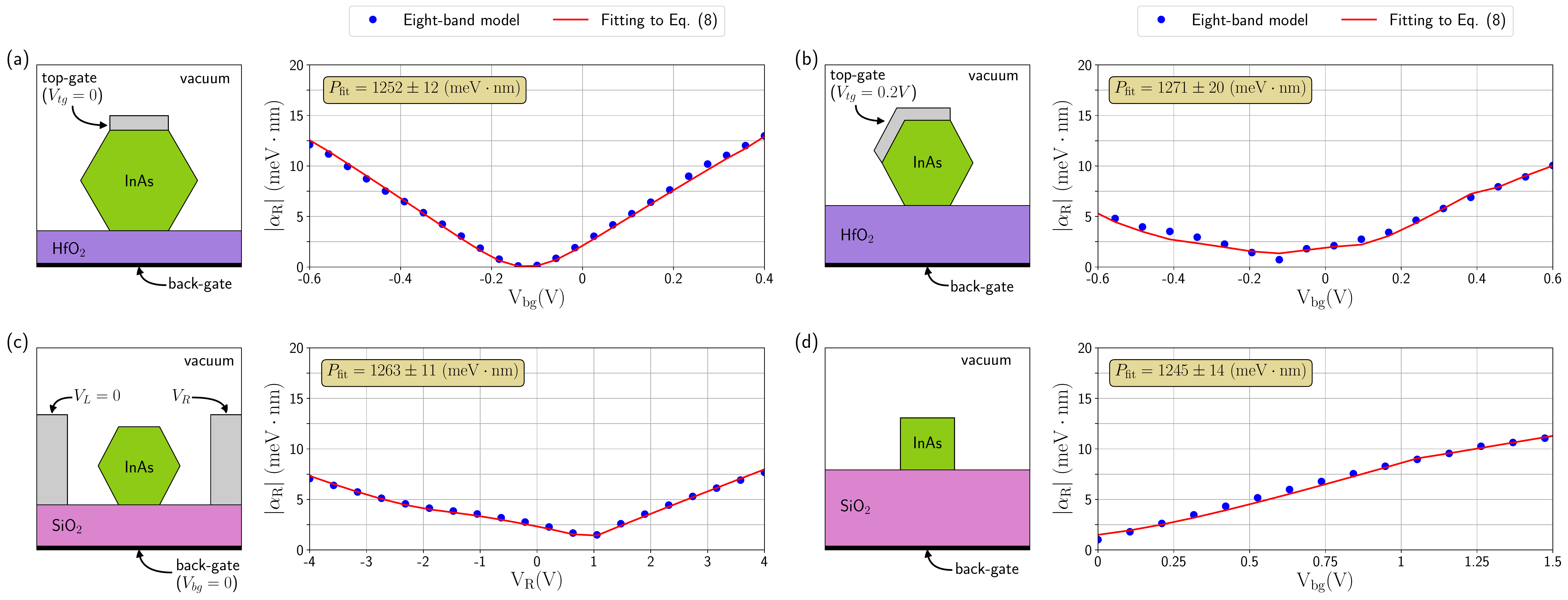}
\caption{Rashba SO coupling modulus versus gate voltage for a Zinc-blende (111) InAs nanowire with different geometries and located in different electrostatic environments, sketched to the left of each sub-figure. Dots correspond to $\alpha_R$ of the lowest-energy subband obtained from the 8B model Hamiltonian using Eq. \eqref{E_fit_kp}. Solid red curves correspond to the conduction band SO coupling obtained from the improved Eq. \eqref{Eq:alpha_R_improved}, by fitting $P_{\mathrm{fit}}$ in each case to the 8B model result. The resulting $P_{\mathrm{fit}}$ values are shown in each sub-figure legend. Parameters: (a) same as in Fig. \ref{Fig2} of the main text; (b) $W_{\rm{wire}}=80$ nm, $W_{\rm{substrate}}=50$ nm, $W_{\rm{layer}}=10$ nm, $\epsilon_{\rm{wire}}=\epsilon_{\rm{InAs}}=15.15$, $\epsilon_{\rm{substrate}}=\epsilon_{\rm{HfO_2}}=25$, $V_{\rm{layer}}=0.2$ V and $\rho_{\rm{surf}}=0$; (c) $W_{\rm{wire}}=60$ nm, $W_{\rm{substrate}}=400$ nm, the separation between side gates and the nanowire is $70$ nm, $\epsilon_{\rm{wire}}=\epsilon_{\rm{InAs}}=15.15$, $\epsilon_{\rm{substrate}}=\epsilon_{\rm{SiO_2}}=3.9$, $V_{\rm{bg}}=0$ V,  $V_{\rm{L}}=0$ V and $\rho_{\rm{surf}}=5\cdot 10^{-3}\left(\frac{e}{nm^3}\right)$; (d) $W_{\rm{wire}}=40$ nm, $W_{\rm{substrate}}=40$ nm, $\epsilon_{\rm{wire}}=\epsilon_{\rm{InAs}}=15.15$, $\epsilon_{\rm{substrate}}=\epsilon_{\rm{SiO_2}}=3.9$ and $\rho_{\rm{surf}}=5\cdot 10^{-3}\left(\frac{e}{nm^3}\right)$. Notice that in this last case the nanowire has a square section instead of hexagonal. The charge density inside the wire $\rho_{\rm{mobile}}$ has been neglected for simplicity in (a-c), while it has been included in the Thomas-Fermi approximation in (d). The 8B model parameters are given in Table \ref{Table2}.}
\label{FigG}
\end{figure}

One of the core assumptions of our work is that the improved Kane parameter $P_{\mathrm{fit}}$ depends weakly on the electrostatic potential. This allows us to use the improved SO coupling equation [Eq. \eqref{Eq:alpha_R_improved}] regardless of the precise electrostatic environment surrounding the wire. To illustrate further this point, we follow the same procedure as the one explained in Sec. \ref{Sec-Results} to extract $P_{\mathrm{fit}}$ from four different devices/environments. Their sketches are depicted to the left of each sub-figure in Fig. \ref{FigG}. To the right, we show the SO coupling of the lowest-energy subband extracted from 8B model simulations (blue dots) versus gate voltage, taking into account the electrostatic environment of its corresponding sketch. We obtain $P_{\mathrm{fit}}$ by fitting Eq. \eqref{Eq:alpha_R_improved} to these values. The fitting curves are shown with solid red lines, and the resulting $P_{\mathrm{fit}}$ values are shown in their respective legends. The geometries considered here are significantly different between them. Simulations for (a) and (b) have two gates (back and top ones), but with different applied voltages and covering different number of wire's facets. In (c) we consider a device with three gates (one back-gate and two side-gates), and we explore the dependence of the SO coupling with one of the side gates, similarly to the device of Ref. \onlinecite{Scherubl:PRB16}. Finally, in (d) we consider only one gate (a back one), but we include the charge density of the wire $\rho_\mathrm{mobile}$ in the Thomas-Fermi approximation, as the device of Ref. \onlinecite{Dhara:PRB09}. We also consider different charge accumulation layers $\rho_\mathrm{surf}$ [in (b) is zero], different substrate widths and materials, and a square cross-section for the nanowire in (d). Despite all these differences, the discrepancy of $P_{\mathrm{fit}}$ between these setups is below 2\%. This implies that the corrections of the electrostatic potential to $P_{\mathrm{fit}}$ are small (for the range of gate potentials studied in this work) and that, in practice, we can neglect the dependence of $P_{\rm{fit}}$ with $\phi(\vec{r})$.

\section{Improved $P_{\rm{mfit}}$ parameter for the effective mass of Zinc-blende InAs, InSb, GaAs and GaSb nanowires}
\label{App:G}

In the main text, Sec. \ref{Sec-Applications}, we have stated that the conduction band effective mass is properly described using Eq. \eqref{Eq:m_eff}, given in terms of the original Kane parameter $P$. However, the intra-valence band couplings ignored in the conduction band approximation that are essential for the SO coupling could also have had an impact on the effective mass. To demonstrate that this is actually not the case, we follow the same reasoning as in Sec. \ref{Sec-CB_approx} for the SO coupling, Eq. \eqref{Eq:alpha_R_improved}, and propose an \textit{improved} equation for the effective mass, where the original $P$ parameter is substituted by an improved one, $P_{\rm{mfit}}$,
\begin{eqnarray}
\frac{1}{m_{\mathrm{improved}}^{(j)}}=\frac{1}{m_e}+ \left<\Psi_T^{(j)}\right|\frac{2P_{\mathrm{mfit}}^2}{3\hbar^2}\left(\frac{2}{\Delta_g+e\phi(\vec{r})+E_T^{(j)}} + \frac{1}{\Delta_{\rm{soff}}+\Delta_g+e\phi(\vec{r})+E_T^{(j)}}\right)\left|\Psi_T^{(j)}\right>.
\label{Eq:m_eff_improved}
\end{eqnarray}
Notice that the improved parameter for the effective mass, $P_{\mathrm{mfit}}$, is not the same as the one for the Rashba SO coupling, $P_{\mathrm{fit}}$, in the same way that the quantity between square brackets in Eq. \eqref{Eq:m_eff} is different from the one in Eq. \eqref{Eq:alpha_R}. This can readily be understood by following the conduction band approximation in terms of the Green's function Dyson series explained in App. \ref{App:B}.

\begin{figure}[h]
\includegraphics[scale=0.27]{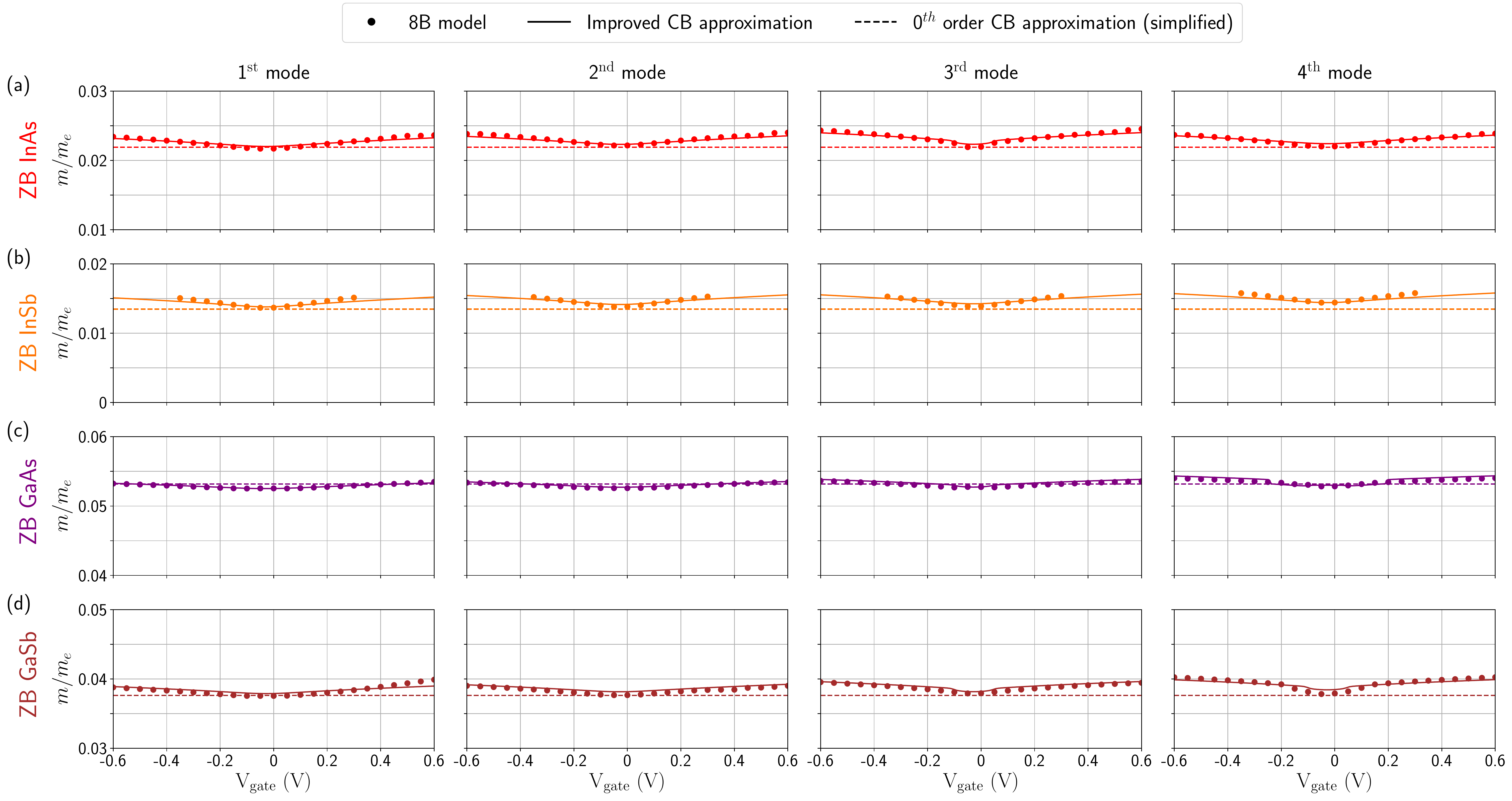}
\caption{Conduction band effective mass $m$ versus gate voltage for different Zinc-blende (ZB) (111) semiconductor nanowires (rows) and different transverse modes (columns). Dots correspond to $m$ obtained from the 8B k$\cdot$p Kane model of App. \ref{App:A} using Eq. \eqref{E_fit_kp}. Solid lines correspond to $m$ obtained from the improved effective mass Eq. \eqref{Eq:m_eff_improved}. The corresponding $P_{\rm{mfit}}$ parameters are collected in Table \ref{Table4}. Dashed lines correspond to the zeroth-order simplified Eq. \eqref{Eq:m_eff-oversimplified}. The simulations have been performed using the same environment and parameters as in Fig. \ref{FigE}.}
\label{FigF}
\end{figure}

Following the same procedure as in Sec. \ref{Sec-Results} for the Rashba SO coupling, we first compute the 8B model band structure for a particular electrostatic environment. We consider here the same one as in Fig. \eqref{FigE}. From the conduction band shape, we can extract the value of the effective mass by fitting Eq. \eqref{E_fit_kp} to the energy subband $j$.  In Fig. \ref{FigF} we show these 8B model results with dots for the first four transverse modes (columns) and for different Zinc-blende (111) semiconductor nanowires (rows). Secondly, we fit Eq. \eqref{Eq:m_eff_improved} to the 8B model results of the lowest-energy subband (first column) in order to get $P_{\rm{mfit}}$ for each material. The resulting fitting parameters are shown in Table \ref{Table4}. We then use these values to compute the improved effective mass with Eq. \eqref{Eq:m_eff_improved}, shown with solid lines in Fig. \ref{FigF}.

Note that the $P_{\rm{mfit}}$ values collected in Table \ref{Table4} turn out to be essentially identical to their corresponding original Kane parameter $P$. This means that the inter-valence band couplings $\gamma_i$ ignored in the zeroth-order conduction band approximation have a minor impact on the effective mass of the conduction band. Only the coupling between conduction and valence bands, accounted by $P$, has a significant contribution to the effective mass. This was further confirmed as a function of the wire's diameter $W_{\rm{wire}}$ in Fig. \ref{FigB}(b) at the end of App. \ref{App:B}. Therefore, we conclude that, even for low-dimensional systems like the finite-width nanowires analyzed here, we can use the original Kane parameter $P$ in the zeroth-order effective mass equation, Eq. \eqref{Eq:m_eff}, to a very good approximation. This is very different from what happens to the SO coupling, as we argue in depth in this paper.

Finally, we point out that the values generally used in the literature for the different semiconductor effective masses correspond to those obtained with the simplified version of the zeroth-order conduction band approximation, Eq. \eqref{Eq:m_eff-oversimplified}. In particular, for the experiments analyzed in Sec. \ref{Sec-Applications} \cite{Dhara:PRB09, Liang:Nano12, Takase:SciRep17, Scherubl:PRB16, Takase:IOP19}, these masses are $m_{\rm{eff}}=0.023m_e$ for InAs and $m_{\rm{eff}}=0.014m_e$ for InSb. In the simplified equation, the assumption $\Delta_g\gg \left|e\phi(\vec{r})+E\right|$ has been made, and thus the effective mass does not depend on the electrostatic potential. Therefore, there could be a deviation between theory and experiments for large gate voltages. To check this, in Fig. \ref{FigF} we show the effective mass provided by the simplified equation with dashed lines. The difference between this approximation and the exact results is always below 5\% for the gate voltage range analyzed here. For larger potential values, the effective mass seems to converge to a constant value, so we expect that the error keeps in the same order of magnitude.

\begin{table}[h]
Table \ref{Table4}: Parameter $P_{\rm{mfit}}$ (in meV$\cdot$nm units) to be used in the improved equation for the effective mass, Eq. \eqref{Eq:m_eff_improved}, within the conduction band approximation. This parameter is extracted by fitting Eq. \eqref{Eq:m_eff_improved} to numerical eight-band model calculations. For comparison, we show the value of the original Kane parameter $P$.
\label{Table4}
\begin{tabularx}{0.6\textwidth} {|>{\centering}X||>{\centering}X|>{\centering\arraybackslash}X|}
\hline 
      &  $P_{\rm{mfit}}$   & $P$        \\ \hline \hline
 InAs (111) &  921$\pm$7  &  919.7   			\\ \hline 
 InSb (111) &  950$\pm$15   &  940.2            \\ \hline
 GaAs (111) &  1039$\pm$5  &  1047.5            \\ \hline
 GaSb (111) &  977$\pm$12  &  971.3             \\ \hline
\end{tabularx}
\end{table}


\end{document}